\documentclass[11pt]{article}
\usepackage{graphicx,dcolumn,bm}
\usepackage{amssymb,amstext,amsmath}
\usepackage{graphicx}
\usepackage{dcolumn}
\usepackage{bm}

\usepackage[top=25.4mm, bottom=25.4mm, left=25.4mm, right=25.4mm]{geometry}

\begin{document}

\title{Cooperation with both synergistic and local interactions can be worse than each alone}
\author{Aming Li$^{1,}$\footnote{liaming@pku.edu.cn} , Bin Wu$^{2,}$\footnote{bin.wu@evolbio.mpg.de} , and Long Wang$^{1,}$\footnote{longwang@pku.edu.cn}}
\date{December 14, 2013}
\maketitle
\begin{enumerate}
 \item Center for Systems and Control, State Key Laboratory for Turbulence and Complex Systems, College of Engineering, Peking University, 100871 Beijing, China
 \item Research Group for Evolutionary Theory, Max-Planck-Institute for Evolutionary Biology,\\ August-Thienemann-Str. 2, 24306 Pl{\"o}n, Germany
\end{enumerate}

%

\begin{abstract}
Cooperation is ubiquitous ranging from multicellular organisms to human societies.
Population structures indicating individuals' limited interaction ranges are crucial to understand this issue.
But it is still at large to what extend multiple interactions involving nonlinearity in payoff play a role on cooperation in structured populations.
Here we show a rule, which determines the emergence and stabilization of cooperation, under multiple discounted, linear, and synergistic interactions.
The rule is validated by simulations in homogenous and heterogenous structured populations.
We find that the more neighbors there are the harder for cooperation to evolve for multiple interactions with linearity and discounting.
For synergistic scenario, however,
distinct from its pairwise counterpart, moderate number of neighbors can be the worst,
indicating that synergistic interactions work with strangers but not with neighbors.
Our results suggest that the combination of different factors which promotes cooperation alone can be worse than that with every single factor.
\end{abstract}

\maketitle


\section{Introduction}

The particulars of  why and how cooperation evolves have perplexed evolutionary biologists and sociologists enduringly \cite{Hamilton63AN,Trivers71QRB,Hofbauer98book,Nowak06book}.
A cooperator takes an altruistic action which supplies a benefit, $b$, for another individual at a cost, $c$,  while a defector does nothing.
One of the main tasks of evolutionary theory is to explain how and why cooperation is present.
Evolutionary game theory provides a powerful platform to understand the evolution of cooperation in unstructured populations, with the replicator equation in
infinite populations \cite{Hofbauer98book} and stochastic dynamics in finite populations \cite{Nowak2004,Traulsen2005,Traulsen2006a}.

Recently the assumption of a well mixed population is removed, and the population allows individuals to interact locally \cite{Nowak92Nature,durrett:TPB:1994}.
Typically networks are adopted to depict such population structure,
since it is simple in definition, while complex in property \cite{Nowak92Nature,Albert2002a}.
The nodes of the network represent individuals, while the edges denote connections in between \cite{Lieberman2005a,Ohtsuki06Nature}.
In this way, a network paves the way to capture the intrinsic idea of local interaction \cite{Gonzalez2008a,Brockmann2006a}: Individuals interact with their neighbors only.
In particular, the degree of a node represents the number of neighbors of the focal individual, which indicates the interaction range.
These network structures are widespread in human organizations \cite{Skyrms2000a,Jackson2000a}, scientific collaboration among researchers \cite{Newman2001a}, and even somatic evolution within multicellular organisms \cite{Nowak2003a}.
However, for structured populations,
in contrast with that in the well mixed case \cite{Nowak2004,Traulsen2005},
it becomes challenging to analyze the evolutionary dynamics theoretically.
This is because enormous possible topological configurations arise during the process of evolution \cite{Barabasi2005aNATPHY,Perc2010,Li2013a}.
In spite of being challenging, there are advances in the analytical methods \cite{Ohtsuki06Nature,Tarnita2009,Antal09PNAS,Nowak10PTRSB,BWu10PLoSONE,BWu11PRE}.
The main result is that local interactions can pave the way for the emergence of cooperation.

The conflict between cooperation and defection is captured by the prisoner's dilemma in the beginning \cite{Trivers71QRB,Luce1957a,Rapoport1965a},
a pairwise game.
Though the ``Tragedy of Commons'', a multi-player game, was introduced to depict this dilemma long before \cite{Hardin1968},
it has not been popular until recently owing to its complexity
\cite{Broom1997a,HAUERT2002,Bach2006,Zheng2007a,Souza2009,Gokhale2010,Pacheco2009,Perc2013a,wu:Games:2013}.
This is also true in structured populations:
For pairwise interactions, conditions for cooperators to be selected over defectors have been theoretically investigated in general structured populations \cite{Ohtsuki06Nature,Tarnita2009,Nowak10PTRSB}.
For multiple interactions, however, only two extreme types of network structure with cycle \cite{Veelen2012a} and well mixed populations \cite{Souza2009,Gokhale2010,Pacheco2009,Kurokawa2009} have been addressed.
For the network degree in between the minimum (the cycle) and the maximum (well-mixed population), it is unclear under what conditions cooperation outperforms defection.
Besides, we introduce nonlinearity in the public goods,
which is intrinsic to the fitness of multi-player games.
For simplicity, the synergistic and discountable effects of the public goods are adopted:
These effects are wide spread in microbes \cite{Hauert2006JTB,Frank2010,Archetti2012}.
As a cluster of microbes secretes enzymes to digest the extracellular resource,
the benefit of the secreted enzyme (public goods) provided by the first cooperative cell may play a vital role for survival,
while the enzymes will eventually be saturated for the resource with the increase of cooperators,
thus the cooperator cells joining the group later only contribute diminishing small benefits to the group \cite{MacLean2010a}.
This is the discounting effect of the public goods.
While for synergy, enzyme-mediated reactions will be launched by enzyme-producing cooperators.
With the concentration of enzyme production, this may exhibit a faster efficiency than linear increase \cite{Hammes1982book}.
In addition, the framework of synergy and discounting effects provides a unifying framework reconciling different social dilemmas \cite{Hauert2006JTB},
thus it does not lose generality in spite of its simplicity.
The main result in the well mixed population is that synergy is beneficial for the emergence of cooperation while the discounting effect is detrimental.

Considering the importance of both the population structure and the multiple interactions on the evolution of cooperation,
we theoretically explore how the combination of these two effects affects the emergence and stabilization  of cooperation.
To this end,
we are addressing the stochastic dynamics of the public goods game with synergy and discounting in a generally random regular graph with arbitrary degree.

The remainder of this paper is organized as follows.
Section \ref{main_2} provides a detail description of the model, and results are given in Section \ref{main_3}.
Conclusion and discussion are drawn in Section \ref{main_4}.
Finally, we present all the theoretical derivations in Appendix \ref{appendix_a} to \ref{appendix_d}.

\section{Model}
\label{main_2}
\begin{figure}
\center
\includegraphics[width=6cm]{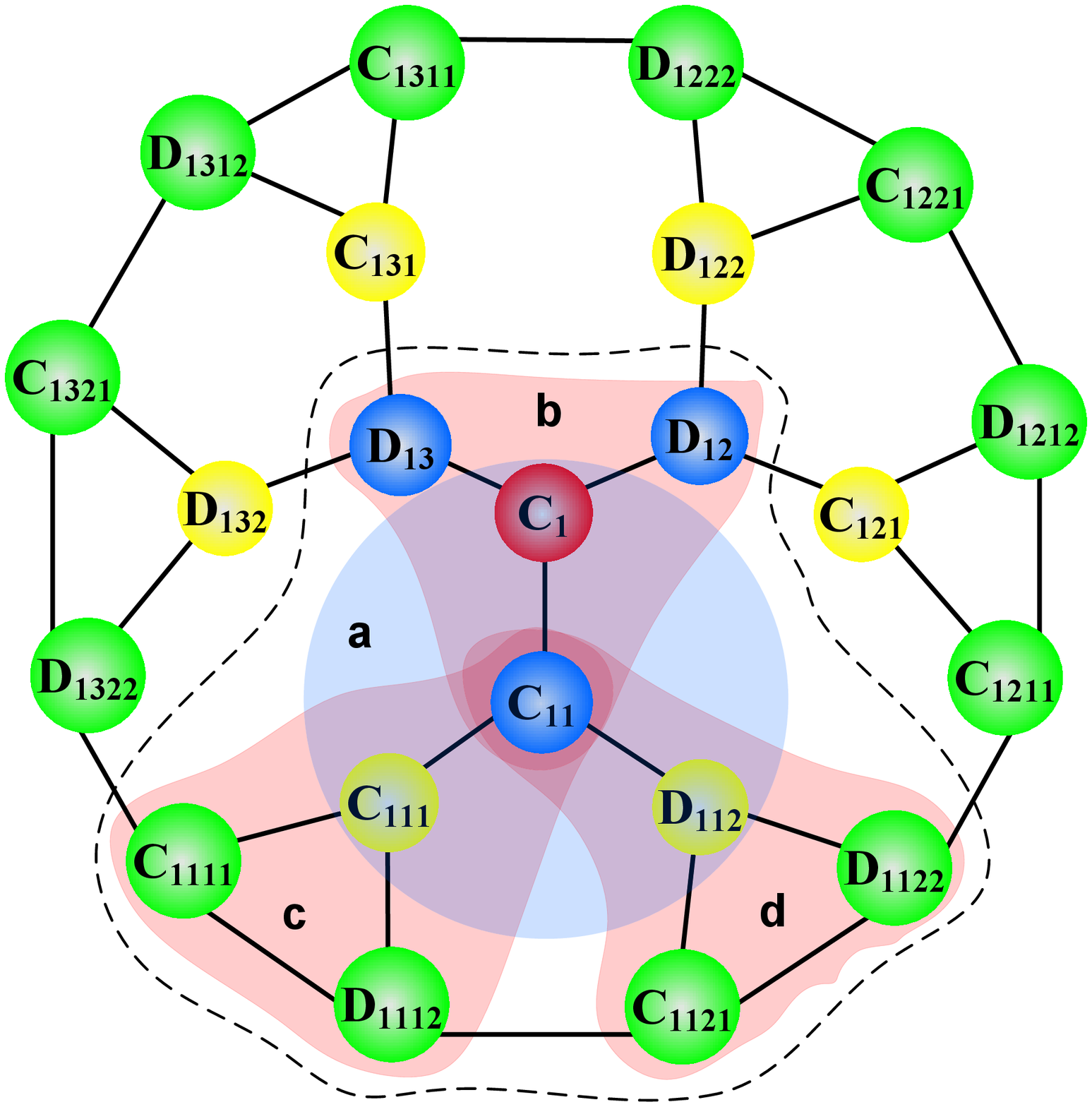}
\caption{\label{fig_1}
(Color online) Illustration of updating on a network.
We show a network with size $N=22$ and every player has $k=3$ neighbors here.
Cooperator $\mathrm{C_{11}}$ and defector $\mathrm{D_{12}}$ ($\mathrm{D_{13}}$) are neighbors of the selected cooperator $\mathrm{C_1}$ with updating.
Both of $\mathrm{C_{11}}$ and $\mathrm{D_{12}}$ ($\mathrm{D_{13}}$) have neighbors with strategy of cooperation or defection except $\mathrm{C_1}$, which are called $\mathrm{C_{111}}$ and $\mathrm{D_{112}}$, $\mathrm{C_{121}}$ ($\mathrm{C_{131}}$) and $\mathrm{D_{122}}$ ($\mathrm{D_{132}}$), respectively.
$\mathrm{C_{1a1}}$ and $\mathrm{D_{1b2}}$ also have neighbors $\mathrm{C_{1a11}}$, $\mathrm{C_{1b21}}$ adopting strategy of cooperation and $\mathrm{D_{1a12}}$, $\mathrm{D_{1b22}}$ with defection, where both of a and b mean 1, 2, or 3.
Each player organizes a public goods game with all of its  $k$ neighbors.
Thus each individual participates in $k+1$ public goods games of size $k+1$ \cite{Santos08Nature}.
As an example, for the payoff of $\mathrm{C_{11}}$,
all players marked within the dashed curve are relevant.
The payoff comes from all games $\mathrm{C_{11}}$ participates in, where one game (shaded in blue) held by $\mathrm{C_{11}}$, that is, part (\textbf{a}), and the other three (shaded in red) held by $\mathrm{C_{1}}$, $\mathrm{C_{111}}$, and $\mathrm{D_{112}}$, that is, part (\textbf{b}), (\textbf{c}), and (\textbf{d}).
}
\end{figure}

We consider a finite population located on a graph of size $N$.
Individuals are assigned to the nodes of the graph, whereas social ties between them are represented by the edges \cite{Lieberman2005a,Ohtsuki06Nature}.
Every individual has $k$ neighbors.
As illustrated in Fig.~\ref{fig_1}, players participate in the public goods game organized by themselves and their neighbors \cite{Santos08Nature}, that is to say, each player participates in $k+1$ public goods games of size $n=k+1$.

For the public goods game, the first cooperator contributes a benefit $b$ while the $j^{\text{th}}$ ($1 \leq j \leq n$) cooperator contributes $b\delta^{j-1}$  to the common pool.
Every cooperator pays the same cost $c$.
Defectors exploit the group by reaping benefits without paying anything.
The accumulated benefits are distributed equally to all the $n$ players in the group irrespective of their behaviors.
Thus, defectors and cooperators receive the following payoffs
\begin{equation}
\begin{split}
  P_D(i) &= \frac{b}{n}(1+\delta+\delta^2+\cdots+\delta^{i-1})=\frac{b}{n}\frac{1-\delta^i}{1-\delta} \\
  P_C(i) &= P_D(i)-c
\end{split}
\end{equation}
where $i$ is the number of cooperators within the group.
Here $\delta>0$ is regarded as the discounting ($0<\delta<1$) or synergy ($\delta>1$) factor.
As $\delta=1$,
it degenerates to the linear public goods game
with $P_D(i)=rci/n$,
where $r=b/c$ is the multiplication factor.

After playing the public goods game, the payoff $P$ of every player is transformed into fitness $f$ by fitness mapping \cite{Nowak2004,BWu10PRE}.
Here we adopt the linear fitness which consists of baseline fitness and the payoffs arising from games \cite{Nowak2004,Ohtsuki06Nature},
i.e., $f=1-w+wP$ where $w$ varying from $0$ to $1$ is the intensity of selection.
For $w\rightarrow0$, the selection is weak.
It means that the game is merely one of many factors which contribute to the entire fitness of an individual \cite{Nowak2004,Ohtsuki06Nature}.

As to the updating rule, the ``death-birth'' (DB) process \cite{Nowak06book} is employed.
Within the process, a player in a population is randomly selected to die at each time step, and then all neighbors of the focused player, with probability proportional to their individual fitness, compete for the vacant site.

We study the emergence of cooperation by comparing the fixation probability \cite{Nowak2004,Ohtsuki06Nature,wu:Games:2013}
of a single cooperator ($\rho_C$) invading a wild population of defective type under weak selection with that under neutrality $1/N$ \cite{Nowak06book}.
If $\rho_C>1/N$ then natural selection favors cooperator replacing defector \cite{Nowak06book}, so we see that natural selection favors the emergence of cooperation.
We see that natural selection favors the stabilization of cooperation if $\rho_D<1/N$, that is, natural selection opposes the fixation of defectors.
And if $\rho_C>\rho_D$, we see that natural selection favors cooperator over defector \cite{Nowak06book}.

\begin{figure*}
\center
\includegraphics[width=15cm]{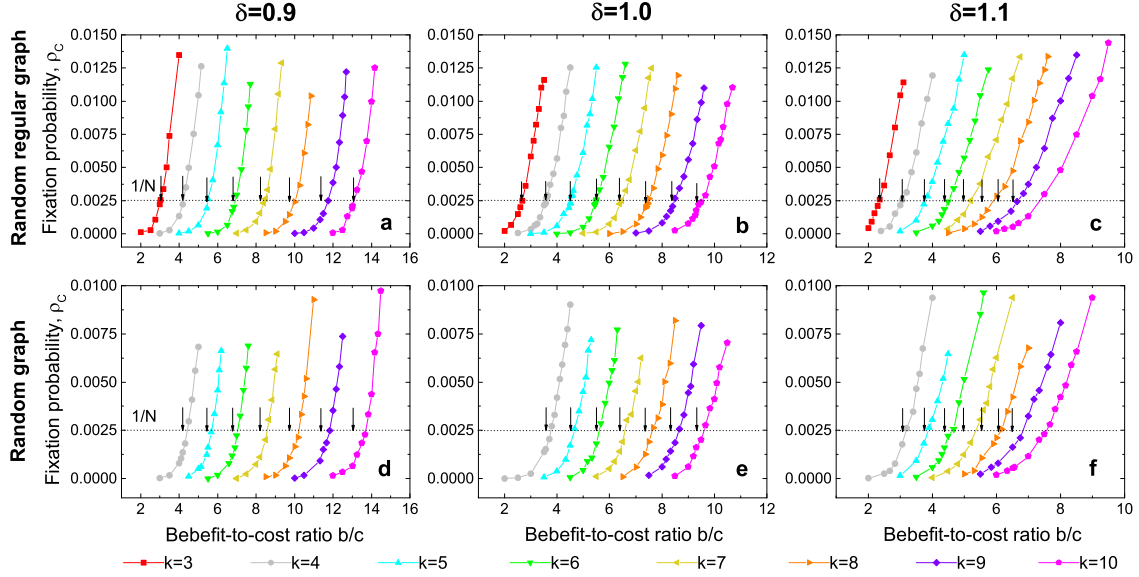}
\caption{\label{fig_2}
(Color online) The rule is in good agreement with numerical simulations.
The upper and lower rows show the simulation results of the fixation probability for the random regular graph with degree $k$ \textbf{(a}, \textbf{b}, and \textbf{c)} and the random graph with average degree $k$ \textbf{(d}, \textbf{e}, and \textbf{f)}.
Both are performed for population size $N=400$ with weak selection $w=0.01$.
Fixation probability $\rho_C$ is approximated by $10^6$ independent runs.
The horizontal dotted line in each figure represents the fixation probability of a neutral mutant, $1/N$.
And the arrows point to the theoretical critical value of $b/c$ favoring cooperation, i.e., $\rho_C>1/N$.
The first to third columns correspond to the public goods game with discounting $\delta=0.9$, linearity $\delta=1.0$, and synergy $\delta=1.1$.
We find that the rule applies to both random regular graphs and random graphs.
Note that, as $\delta$ is too big or small, accumulation of the payoff in the common pool is changing rapidly.
Any extremely big or small contribution $b\delta^{j-1}$ of any later $j^\text{th}$ cooperator are fabricated, thus only $0.9\leq\delta\leq1.1$  are considered here.
The discrepancy is larger for the random graph \textbf{(f)} with high average degree ($k=10$) since the derivation of the rule is at large $N$ and pair approximation is formulated for the regular graph without any loops.
}
\end{figure*}

\section{Results}
\label{main_3}

We obtain the fixation probability of both cooperation and defection by the pair approximation (see Appendix \ref{appendix_a} and \ref{appendix_b}).
For large population size and weak selection, we have a rule: $\rho_C>\frac{1}{N}$ if and only if
\begin{equation}
\label{mainphoccc}
 b/c>
 \begin{cases}
   \frac{(k+1)^2}{k+3}
      & \text{for $\delta=1$}  \\ \\
  \frac{18}{3\delta^2+4\delta+3}
      & \text{for $k=2$}  \\ \\
  \frac{(N-1)(k+1)Q_1}{2Nf(\delta)}
      & \text{otherwise}
 \end{cases},
\end{equation}
and $\rho_D<\frac{1}{N}$ if and only if
\begin{equation}
\label{mainphoddd}
 b/c>
 \begin{cases}
   \frac{(k+1)^2}{k+3}
      & \text{for $\delta=1$}  \\ \\
  \frac{18}{3\delta^2+4\delta+3}
      & \text{for $k=2$}  \\ \\
  \frac{(N-1)(k+1)Q_1}{2Ng(\delta)}
       & \text{otherwise}
 \end{cases}
\end{equation}
where the notations $Q_1$, $f(\delta)$, and $g(\delta)$ can be found in Appendix \ref{main_thesimplerule}.

\begin{figure*}
\center
\includegraphics[width=15cm]{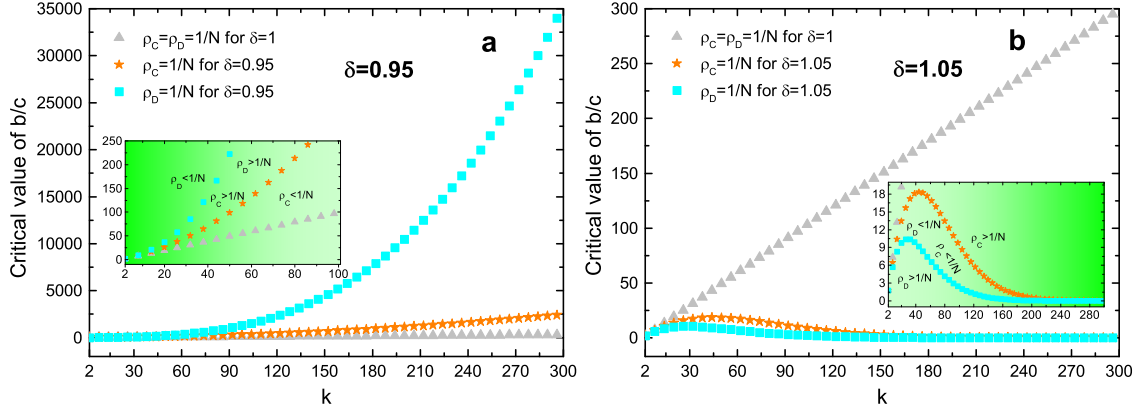}
\caption{\label{fig_3}
(Color online) The critical value of the benefit-to-cost ratio for natural selection favoring emergence and stabilization of cooperation (defection).
We set $\delta=0.95$ \textbf{(a)} and $1.05$  \textbf{(b)} to represent respectively the weak discounting and synergy in structured populations with $N=5\times 10^3$ numerically.
For discounting effect $\delta<1$, both the critical values for $\rho_C=1/N$ and $\rho_D=1/N$ increase rapidly with average degree $k$.
Yet the critical value for $\rho_D=1/N$ is greater than that of $\rho_C=1/N$, and it also increases much faster.
This shows that in the discounting public goods game, with the increase in the number of neighbors, it is easier for a cooperator to be invasive ($\rho_C>1/N$) than to be stabilized ($\rho_D<1/N$).
For synergy effect $\delta>1$, the critical value for $\rho_C=1/N$ is greater than that of $\rho_D=1/N$ for any neighbor size.
This shows that it is easier for cooperation to be stabilized  ($\rho_D<1/N$) than to be invasive ($\rho_C>1/N$).
Interestingly, both the critical values for $\rho_C=1/N$ and $\rho_D=1/N$ are non-monotonic and is a one-hump function of $k$.
This shows for both the emergence and the stabilization of cooperation in the public goods game with weak synergy, a moderate number of neighbors is the worst.
}
\end{figure*}

For the linear public goods game ($\delta=1$), natural selection favors not only the emergence of cooperation but also its stabilization if and only if
\begin{equation}
\label{eq:1:b/c}
b/c>(k+1)^2/(k+3).
\end{equation}
On this occasion, the rule is equivalent to $r>n^2/(n+2)$ where $r$ is the multiplication factor of the common pool.
Since $n^2/(n+2)<n$,
it implies theoretically that cooperative dilemma can be relaxed in structured populations compared with that in the well mixed case,
without invoking any other additional mechanisms.

We get $\rho_C>\rho_D$ if and only if $b/c>9/5$ for linear public goods game on the cycle ($k=2$ in our model).
And the same result is also found in \cite{Veelen2012a}, where, using a direct approach without approximations, van Veelen \textit{et al.}  pointed that $\rho_C>\rho_D$  if and only if $b/c>2n/(4-2/n)$.
As group size $n=3$, the linear public goods game on cycle coincides with that in the structured population with $k=2$, and both deduce $b/c>9/5$.
What's more is that the general rule to determine the emergence of cooperation is found to be in good agreement with computer numerical simulations (see the first row of Fig.~\ref{fig_2}).
And it also approximately applies to heterogeneous structured populations (see the second row of Fig.~\ref{fig_2}).

Based on the equation~\eqref{eq:1:b/c}, we find that $\rho_C>\rho_D$ is also equivalent with the emergence $\rho_C>1/N$ and stabilization $\rho_D<1/N$ of cooperation.
Furthermore, the following equivalence holds
\begin{eqnarray}
\label{pcpd1n}
  \rho_C>\frac{1}{N} \Leftrightarrow \rho_C>\rho_D  \Leftrightarrow \rho_D<\frac{1}{N}
\end{eqnarray}
for the public goods game which is either linear  $\delta=1$ or on a cycle $k=2$.
That is to say, for large structured population under linear public goods game or public goods game with nonlinearity in individual payoff on cycle, we have that natural selection favors emergence of cooperation if and only if it favors stabilization of cooperation.
Further,
we show that the critical value, both for the emergence and the stabilization of cooperation, is continuous with the discounting or synergy factor $\delta$ (see Appendix \ref{appendix_c1}).
Hence the equivalent proposition (\ref{pcpd1n}) applies for infinitesimal nonlinearity (see Appendix \ref{appendix_c2}).

\begin{figure*}
\center
\includegraphics[width=15cm]{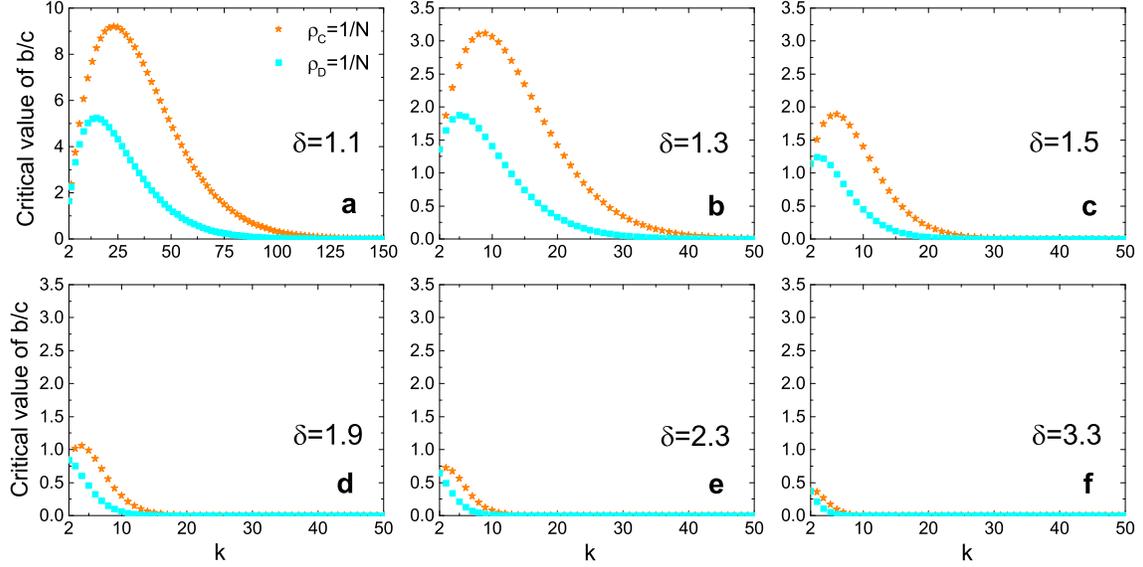}
\caption{\label{fig_4}
(Color online) Critical value of benefit-to-cost ratio degenerates to a monotonic function of average degree $k$ from one-hump function with the enhancement of synergy.
We set different synergy factors to investigate its effects on the critical value of $b/c$ with population size $N=5\times 10^3$.
Specific values of $\delta$ are marked on each panel.
Orange star and cyan square are used to indicate $\rho_C=1/N$ and $\rho_D=1/N$ respectively.
The critical value decreases with the increase of $\delta$.
And it is a parabolic function of $k$ when synergy is weak (\textbf{a}, \textbf{b}, \textbf{c}, \textbf{d}, and \textbf{e}), whereas monotonic when synergy is strong (\textbf{f}).
Under the synergistic effect, big $k$ or $\delta$ can drive the critical value to approach $0$.
The intuitive understanding behind this is the competition between the two factors.
For large $k$, the population is approximately well mixed, thus the local interaction diminishes by the replicator equation \cite{Hauert2006JTB}, and cooperation with synergy thrives, leading to a decrease of the critical value with increasing $k$.
For small $k$, the local interaction plays an important role if the synergy effect is not strong.
That is to say for $\delta$ slightly more than the unit, increasing $k$ does inhibit the fixation for both strategies as
in the pairwise Prisoner's Dilemma \cite{Ohtsuki06Nature}.
To sum up, for small $\delta>1$, there is a hump for the critical value with the increase of neighbors.
For large $\delta$, however, even for small $k$, the synergy effect is strong enough to outperform the locality of the population structure.
By the same argument, the critical value is monotonically decreasing with the neighbors.
}
\end{figure*}

From the rule (see inequalities \eqref{mainphoccc} and \eqref{mainphoddd}), we theoretically get the critical benefit-to-cost ratio $b/c$ for the emergence and stabilization of cooperation (defection) with the two factors combined, saying the spatial reciprocity and nonlinearity in payoff induced by multiple interactions.
Fig.~\ref{fig_3} shows that
weak discounting significantly inhibits both the emergence and stabilization of cooperation,
whereas the weak synergy favors them greatly.
In contrast with the linear public goods game,
the critical ratios $b/c$ for $\rho_C=1/N$ and $\rho_D=1/N$ are no longer overlapping for nonlinear payoff effects (see Fig.~\ref{fig_3}).
In other words, taking into account either discounting or synergy, a particular form of nonlinearity in payoff, the emergence and the stabilization of cooperation are no longer equivalent as in the linear public goods case  (see equivalent proposition (\ref{pcpd1n})).

For multiple interactions such as linear public goods, similar to pairwise interactions \cite{Ohtsuki06Nature}, cooperation will also be impeded with an increase of the number of  neighbors (see light gray up triangle in Fig.~\ref{fig_3}).
Discounting in payoff significantly inhibits cooperation (see Fig.~\ref{fig_3}a).
In particular, in this case, with the increase in the size of neighbourhood,
it will become even harder for the emergence and stabilization of cooperation.
The critical benefit-to-cost ratio is increasing much rapidly than its linear public goods game counterpart.

For weak synergy, the critical benefit-to-cost ratio $b/c$ for emergence or stabilization of cooperation still first increases as the growth of every player's number of neighbors (see Fig.~\ref{fig_3}b).
But it decreases as neighbor size is big enough and tends towards zero.
This illustrates that for small size of neighbor, increasing interaction range, i.e. $k$, is detrimental for cooperation, which is consistent with linear public goods game;
yet the interesting story comes along when the interaction range is relatively large,
in this case, increasing the interaction range is beneficial for cooperation,
which is seldom observed in cooperative dilemmas.
An intuitive understanding can be:
for small neighbor size, the local competition plays a key role.
Even though the public goods are exponentially increasing with additional cooperator in a group, the group is small in size generally, thus the defector would reproduce more efficiently with the increase of size.
This leads to the increase of the critical benefit-to-cost ratio.
For large size of neighbors, however, the exponential increase in the accumulation of public goods with an additional cooperator outperforms the reproduction of defectors.
For example, with the $j^\text{th}$ ($0\leq j \leq k+1$) cooperator's benefit $b\delta^{j-1}$ to common pool,  big $j$ induces large payoff $b\delta^{j-1}/(k+1)$ to every player in the same group with weak synergistically enhanced effect $\delta$ as well as the baseline benefit $b$.
This rather large payoff paves the way for emergence and stabilization of cooperation for large neighbor size.
Therefore, it is the worst for emergence of cooperation when the number of neighbors is moderate.
Another intuitive explanation is that for weak synergy effect, the replicator dynamics allows the coexistence of cooperators and defectors \cite{Hauert2006JTB}, which is quite similar to the snowdrift game.
For the snowdrift game, however, local interaction can inhibit cooperation \cite{Hauert04Nature} in contrast with the prisoner's dilemma \cite{taylor:TPB:2006}.
Here, we explicitly show up to how many numbers would be the worst for cooperation in such scenario.
It shows that synergistic interactions work with strangers in well mixed populations \cite{Hauert2006JTB}.
Furthermore, we also find that the synergistic interactions do not work with neighbors in structured populations.

For stronger synergy,
the critical benefit-to-cost ratio always decreases with the neighbor size (see Fig.~\ref{fig_4}).
Thus in this case cooperation could be promoted significantly with the increase of every player's number of neighbors under strong synergy, which intrinsically differs from effect of linearity, discounting, or weak synergy.
This is because the synergy effect is so great that it effectively is a coexistence game where best relies should be of the population minority.
In this case, enlarging the interaction range
paves the way for cooperator mutants to be more like an minority yielding an enhancement of cooperation level.
Our results strongly suggest that the intrinsic multiple interactions, where payoffs are nonlinear in general, can cause the so called sea-sickness \cite{hamilton:bookchapter:1975}.

\section{Conclusion and Discussion}
\label{main_4}
In conclusion, we find a rule theoretically elucidating the critical value of benefit-to-cost ratio $b/c$ up to which cooperation emerges and is stabilized.
In addition numerical simulations verify the validity of the rule as well as its feasibility for random graph.
For linear public goods game on any regular graph
and any public goods game with synergy and discounting on a cycle,
we find an equivalent proposition that the rule determines not only the emergence, but also the stabilization of cooperation.
What's more, in public goods with synergy, we present that it can be the worst for the emergence of cooperation, as the number of neighbors is moderate.
We find that synergistic interactions work with strangers but not with neighbors,
and cooperation with both synergistic and local interactions can be worse than that with each alone.
Our work suggests that there can be a big shadow in the effects of combinational mechanisms on the evolution of cooperation.

Over the past two decades, structured populations depicted by networks have been taken into consideration to study the evolution of cooperation by virtue of evolutionary graph theory\cite{Lieberman2005a,Perc2010,Nowak10PTRSB,Szabo07review}.
It has been shown that cooperation can flourish in both static network and dynamic network \cite{Nowak92Nature,Ohtsuki06Nature,Perc2010,BWu10PLoSONE,Szabo07review,Szabo02PRL,Pacheco2006}
 (for an exception, please see \cite{Hauert04Nature}).
The main reason is that these population structures can lead to a clustering of cooperative individuals \cite{Nowak92Nature,Ohtsuki06Nature,Li2013a,Nowak10PTRSB,Fletcher09PRSB},
within which cooperators can survive by enjoying the benefits from mutual cooperation even though some cooperators are exploited by defectors along cluster boundaries.
We indicate that in a non-additive public goods game,
where nonlinearity in payoff arises,
this clustering (see equations (\ref{appendix_nete1})  and (\ref{appendix_nete2}) in Appendix \ref{appendix_a2}) is not always beneficial when the neighbors are few in number:
in a synergy public goods scenario, where the latter cooperators in the group contribute significantly much more than the previous cooperators, the worst case for cooperation emerges when the number of neighbors is moderate, not too big nor too small.
This means that synergistic interactions work with strangers but not with neighbors.
Our results show that the interaction between different mechanisms \cite{Veelen2012b} might trigger novel unexpected results.
The combination of different factors with each promoting cooperation alone can be worse than every factor alone in promoting cooperation.
Thus, it may be promising to investigate the combination of previous mechanisms promoting cooperation.

We find that the rules governing the emergence and stabilization of cooperation are equivalent for linear public goods games, which is validated by numerical simulations on homogenous as well as heterogenous structured populations.
The rule simply asks the benefit to cost ratio $b/c$ to exceed a critical value $(k+1)^2/(k+3)$,
where $k$ is the average number of neighbors in the population.
In fact, for any number of neighbors $k$, the numerator of the critical value is
$(k+1)^2$, which suggests the number of individuals relevant to the payoff of the focal individual with recounting (see Fig.~\ref{fig_1}),
i.e., the product of the group size $k+1$ and the average number of the public goods game every player involved in, $k+1$.
Therefore, in this case, as in a well mixed population, multiple interactions significantly inhibit the cooperation than its pairwise counterpart.
For public goods games on a cycle, with either linearity, discounting, or synergy.
The equivalence still holds between the rules facilitating the emergence of cooperation and that governing the stabilization.
Therefore the same criterion applies to determine under what condition the average abundance of cooperation exceeds that of defection in the mutation-selection equilibrium under small mutation \cite{fudenberg:JET:2006,wu:JMB:2012}.

The equivalence falls down for general population structures and nonlinear public goods game.
For the synergy effect, the emergence and the stabilization of cooperation are facilitated significantly for any number of neighbors compared with the linear public goods game.
Being stabilized, in this case, is much easier than the emergence.
For the discounting effect, both the emergence and the stabilization of cooperation are inhibited  significantly for any number of neighbors compared with the linear public goods game.
Being stabilized, in this case, is much harder than the emergence.
Therefore, both synergy and discounting has a more significant role in the stabilization compared with the emergence.

As applications, for microbes with either synergy and discounting public goods, in particular,
if the public goods are diffusive \cite{Gore2009a}, the average degree of the network suggests the diffusion rate of the public goods.
Our result suggests that for the discounting public goods game,
only a low diffusion rate of public goods can make the cooperator cells thrive;
For the synergy public goods game, however, it is always better than the discounting case.
The interesting result lies in the fact that cooperation is better off for both very low and very high diffusion, whereas is worst off for moderate diffusion rate.
Experimental validation along this line might be interesting.

\section*{Author Contributions}
A.L., B.W. and L.W. devised and analysed the model. A.L. performed theoretical derivation and numerical simulations. A.L., B.W. and L.W. analysed the results and wrote the paper.

\section*{Acknowledgements}
A. Li and L. Wang are supported by NSFC (Grants No. 61375120 and No. 61020106005).
B. Wu greatly acknowledges the sponsorship by the Max-Planck Society.
We are indebted to Yong Wang, Maozheng Guo, Guangming Xie, James Price and Arne Traulsen for helpful discussions.
Discussions with Xi Weng and Te Wu are also acknowledged.

\newpage
\appendix
\section{The calculation of fixation probability under the framework of pair approximation}
\label{appendix_a}
We adopt the fixation probability to investigate the emergence and stabilization of cooperation, which has been used by many researchers for finite population
\cite{Ewens2004a,Nowak2004,Ohtsuki06Nature,CrowandKimura2009}.
The updating process and dynamics can be got by pair approximation \cite{Ohtsuki06Nature,MATSUDA1992a}.
And the method to derive fixation probability of cooperation or defection based on multiple interactions is similar to that based on pairwise interactions \cite{Ohtsuki06Nature}.
\subsection{Updating process}
For the structured population, let $X$ and $Y$ denote the strategy of cooperation or defection.
And $p_X$ and $p_{XY}$ are assigned respectively to the frequency of strategy $X$ and $XY$ pairs.
$q_{X|Y}$ indicates the conditional probability of finding a player of strategy $X$ for a player of strategy $Y$.
From the notations interpreted above, we have
\begin{equation*}
\begin{split}
  p_C+p_D &= 1, \\
  q_{C|X}+q_{D|X} &= 1, \\
  p_{XY} &= p_Y \cdot q_{X|Y}, \\
  p_{CD} &= p_{DC}.
\end{split}
\end{equation*}
Furthermore, under the framework of pair approximation, we find $p_C$ and $q_{C|C}$ are sufficient
to describe the system since
\begin{eqnarray}
  p_D &=& 1-p_C,  \nonumber \\
  \label{basic1}
  q_{D|C} &=& 1-q_{C|C}, \\
  \label{basic2}
  p_{CD} &=& p_{DC} = p_C \cdot q_{D|C}  = p_C(1-q_{C|C}), \\
  \label{basic3}
  q_{C|D} &=& \frac{p_{CD}}{p_D} = \frac{p_C(1-q_{C|C})}{1-p_C}, \\
 \label{basic4}
  q_{D|D} &=& 1-q_{C|D} = \frac{1-2p_C+p_Cq_{C|C}}{1-p_C}, \\
   \label{basic5}
  p_{DD} &=& p_Dq_{D|D} = 1-2p_C+p_Cq_{C|C}.
\end{eqnarray}

We assume that a selected individual, cooperator $\mathrm{C_1}$, who has $k$ neighbors which consist of $k_C$ cooperators ($\mathrm{C_{11}}$) and $k_D$ defectors ($\mathrm{D_{12}}$) (see Fig.~$1$ in main text), is replaced by a defector with probability $\Lambda$ under the death-birth (DB) process, thus we have
\begin{equation*}
\begin{split}
\Lambda
&= \frac{k_D(1-w+w\pi_D^C)}{k_C(1-w+w\pi_C^C)+k_D(1-w+w\pi_D^C)} \\
&= \frac{k_D}{k}+\frac{k_Ck_D(\pi_D^C-\pi_C^C)}{k^2}w+O(w^2)
\end{split}
\end{equation*}
with $k_C+k_D=k$, where $\pi^D_C$, $\pi^D_D$ mean the payoff of $\mathrm{C_{11}}$, $\mathrm{D_{12}}$ after playing public goods game respectively,
and $O(w^i)$ means that the error is of order $O(w^i)$, where $i$ is a positive integer.

In order to compute the payoff of an individual after playing public goods games, we should consider the population structure around the individual.
Here, we list the the neighbor and the number of cooperators as well as defectors among $k$ neighbors of $\mathrm{X_{ab}}$ and her neighbors as
\begin{equation*}
\label{table1}
\mathrm{X_{ab}}
\left\{
  \begin{array}{lll}
         \mathrm{X_a}: 1
            \left\{
            \begin{array}{lll}
            \mathrm{C_{a1}}: k_C \\
            \mathrm{D_{a2}} \text{~and~} \mathrm{D_{a3}}: k_D
            \end{array}
            \right. \vspace{0.3cm}
         \\
         \begin{array}{ll}
         \mathrm{C_{ab1}}: (k-1)q_{C|X}
         \end{array}
            \left\{
            \begin{array}{lll}
            \mathrm{X_{ab}}: 1 \\
            \mathrm{C_{ab11}}: (k-1)q_{C|C} \\
            \mathrm{D_{ab12}}: (k-1)q_{D|C}
            \end{array}
            \right. \vspace{0.3cm}
         \\
         \begin{array}{ll}
         \mathrm{D_{ab2}}: (k-1)q_{D|X}
         \end{array}
            \left\{
            \begin{array}{lll}
               \mathrm{X_{ab}}: 1 \\
               \mathrm{C_{ab21}}: (k-1)q_{C|D} \\
               \mathrm{D_{ab22}}: (k-1)q_{D|D}
            \end{array}
            \right.
  \end{array}
\right.
\end{equation*}
where $\mathrm{X}$ can be $\mathrm{C}$ or $\mathrm{D}$, $a=1$ or $2$, and $b=1$, $2$, or $3$.

Therefore, we know that there are $(k-1)q_{C|C}$ cooperators (we call them $\mathrm{C_{111}}$), except $\mathrm{C_{1}}$, and $(k-1)q_{D|C}$ defectors (we call them $\mathrm{D_{112}}$) among the $k$ neighbors of $\mathrm{C_{11}}$ (see Fig.~$1$ in main text).
Hence, we have
\begin{eqnarray*}
 \pi^C_C&=&P_C\left((k-1)q_{C|C}+2\right)+(k-1)q_{C|C}\cdot P_C\left((k-1)q_{C|C}+2\right)
 \nonumber \\ & &
 +(k-1)q_{D|C}\cdot P_C\left((k-1)q_{C|D}+1\right)+P_C(k_C+1)
\end{eqnarray*}
considering population structure around $\mathrm{C_{11}}$.
The first term in the right-hand side of the above equation is the payoff got by $\mathrm{C_{11}}$ from the public goods game centered by itself, and the last three terms are the sum of the payoffs from the games organized by its neighbors.
Similarly, among the $k$ neighbors of $\mathrm{D_{12}}$ (see Fig.~$1$ in main text), there are $(k-1)q_{C|D}$ cooperators (we call them $\mathrm{C_{121}}$), except $\mathrm{C_{1}}$, and $(k-1)q_{D|D}$ defectors (we call them $\mathrm{D_{122}}$).
Thus we have
\begin{eqnarray*}
 \pi^C_D&=&P_D\left((k-1)q_{C|D}+1\right)+(k-1)q_{C|D}\cdot P_D\left((k-1)q_{C|C}+1\right)
 \nonumber \\ & &
 +(k-1)q_{D|D}\cdot P_D\left((k-1)q_{C|D}\right)+P_D(k_C+1)
\end{eqnarray*}
considering population structure around $\mathrm{D_{12}}$.
From the point of pair approximation, we know that $\mathrm{D_{13}}$ can obtain the same payoff as $\mathrm{D_{12}}$.

If the selected cooperator is replaced by a defector, $p_C$ will decreases by $\frac{1}{N}$ with probability
\begin{eqnarray*}
  \mathrm{Pr}\Big(\Delta p_C=-\frac{1}{N}\Big)
   &=&
   p_C\sum_{k_C=0}^{k}{k \choose k_C}q_{C|C}^{k_C}q_{D|C}^{k_D}\Lambda.
\end{eqnarray*}

Simultaneously, as the decrease of $p_C$, the number of $CC$-pairs is changing.
And $CC$-pairs will decrease by $k_C$ after a defector replacing a cooperator.
Hence $p_{CC}$ decreases by $\frac{k_C}{kN/2}$ with probability
\begin{equation*}
\mathrm{Pr}\Big(\Delta p_{CC} = -\frac{2k_C}{kN}\Big) = p_C{k \choose k_C}q_{C|C}^{k_C}q_{D|C}^{k_D}\Lambda.
\end{equation*}

If the selected individual is a defector $\mathrm{D_2}$, we obtain
\begin{eqnarray*}
 \pi^D_C  &=&
 P_C\left((k-1)q_{C|C} +1\right)+P_C(k_C)
  \nonumber \\ & &
 +(k-1)q_{C|C}\cdot P_C\left((k-1)q_{C|C}+2\right)
  \nonumber \\ & &
 +(k-1)q_{D|C}\cdot P_C\left((k-1)q_{C|D}+1\right) \\
  \pi^D_D &=& P_D\left((k-1)q_{C|D}\right)+P_D(k_C)
  \nonumber \\ & &
 +(k-1)q_{C|D}\cdot P_D\left((k-1)q_{C|C}+1\right)
 \nonumber \\ & &
 +(k-1)q_{D|D}\cdot P_D\left((k-1)q_{C|D}\right)
\end{eqnarray*}
where $\pi^D_C$ ($\pi^D_D$) indicates the payoff of $\mathrm{D_2}$'s neighbor $\mathrm{C_{21}}$ ($\mathrm{D_{22}}$ or $\mathrm{D_{23}}$) who adopts the strategy of cooperation (defection) after playing  public goods games.
Thus, $\mathrm{D_2}$ is replaced by a cooperator during the updating process with probability
\begin{equation*}
\begin{split}
\Gamma
&= \frac{k_C(1-w+w\pi_C^D)}{k_C(1-w+w\pi_C^D)+k_D(1-w+w\pi_D^D)}.
\end{split}
\end{equation*}
And, similarly we have
\begin{eqnarray*}
  \mathrm{Pr}\Big(\Delta p_C = \frac{1}{N}\Big)
   &=&
   p_D\sum_{k_C=0}^{k}{k \choose k_C}q_{C|D}^{k_C}q_{D|D}^{k_D}\Gamma,
\end{eqnarray*}
and
\begin{eqnarray*}
  \mathrm{Pr}\Big(\Delta p_{CC} = \frac{2k_C}{kN}\Big)
  &=&
p_D{k \choose k_C}q_{C|D}^{k_C}q_{D|D}^{k_D}\Gamma.
\end{eqnarray*}

\subsection{Updating dynamics}
\label{appendix_a2}
If we assume every DB updating incident takes place in one unit of time, the derivatives of $\dot{p_C}$ and $\dot{p}_{CC}$ can be written as
\begin{eqnarray}
\label{pc}
  \dot{p_C}
&=&
    \frac{1}{N}\mathrm{Pr}\left(\Delta p_C=\frac{1}{N}\right)+\left(-\frac{1}{N}\right)\mathrm{Pr}\left(\Delta p_C=-\frac{1}{N}\right),
\end{eqnarray}
and
\begin{eqnarray}
\label{pcc}
  \dot{p}_{CC}
  &=&
    \sum_{k_C=0}^{k}\frac{2k_C}{kN}\mathrm{Pr}\left(\Delta p_{CC} = \frac{2k_C}{kN}\right)+\sum_{k_C=0}^{k}\left(-\frac{2k_C}{kN}\right)\mathrm{Pr}\left(\Delta p_{CC} = -\frac{2k_C}{kN}\right).
\end{eqnarray}

We have
\begin{eqnarray}
\label{pcposi}
  \mathrm{Pr}\Big(\Delta p_C = \frac{1}{N}\Big)
&=&
    p_D\sum_{k_C=0}^{k}{k \choose k_C}q_{C|D}^{k_C}q_{D|D}^{k_D}\frac{k_C}{k}
\nonumber \\ & &
    +p_D\sum_{k_C=0}^{k}{k \choose k_C}q_{C|D}^{k_C}q_{D|D}^{k_D}\frac{k_Ck_D(\pi_C^D-\pi_D^D)}{k^2}w+O(w^2)
\nonumber \\ &=&
    p_D\sum_{k_C=1}^{k}\frac{k!}{k_C!(k-k_C)!}\frac{k_C}{k}q_{C|D}^{k_C}q_{D|D}^{k-k_C}
\nonumber \\ & &
    +p_D(\pi_C^D-\pi_D^D)w\sum_{k_C=0}^{k}\frac{k!}{k_C!(k-k_C)!}\frac{k_C(k-k_C)}{k^2}q_{C|D}^{k_C}q_{D|D}^{k-k_C}+O(w^2)
\nonumber \\ &=&
    p_Dq_{C|D}\sum_{k_C=1}^{k}\frac{(k-1)!}{(k_C-1)!(k-k_C)!}q_{C|D}^{k_C-1}q_{D|D}^{k-k_C}+O(w^2)
\nonumber \\ & &
    +p_Dq_{C|D}q_{D|D}\frac{k-1}{k}(\pi_C^D-\pi_D^D)w\sum_{k_C=1}^{k-1}\frac{(k-2)!}{(k_C-1)!(k-k_C-1)!}q_{C|D}^{k_C-1}q_{D|D}^{k-k_C-1}
\nonumber \\ &=&
    p_Dq_{C|D}+p_Dq_{C|D}q_{D|D}\frac{k-1}{k}(\pi_C^D-\pi_D^D)w+O(w^2)
\nonumber \\ &=&
    p_{CD}+\frac{k-1}{k}(\pi_C^D-\pi_D^D)p_{CD}q_{D|D}w+O(w^2),
\end{eqnarray}
and
\begin{eqnarray}
\label{pcneg}
  \mathrm{Pr}\Big(\Delta p_C=-\frac{1}{N}\Big)
   &=&
    p_C\sum_{k_C=0}^{k}{k \choose k_C}q_{C|C}^{k_C}q_{D|C}^{k_D}(1-\frac{k_C}{k})
    \nonumber \\  & &
    +p_C\sum_{k_C=0}^{k}{k \choose k_C}q_{C|C}^{k_C}q_{D|C}^{k_D}\frac{k_Ck_D(\pi_D^C-\pi_C^C)}{k^2}w+O(w^2)
   \nonumber \\  &=&
  p_C- p_C\sum_{k_C=0}^{k}\frac{k!}{k_C!(k-k_C)!}\frac{k_C}{k}q_{C|C}^{k_C}q_{D|C}^{k-k_C}
  \nonumber \\  & &
  +p_C\sum_{k_C=0}^{k}\frac{k!}{k_C!(k-k_C)!}\frac{k_C(k-k_C)}{k^2}q_{C|C}^{k_C}q_{D|C}^{k-k_C}(\pi_D^C-\pi_C^C)w+O(w^2)
       \nonumber \\  &=&
  p_C- p_Cq_{C|C}\sum_{k_C=1}^{k}\frac{(k-1)!}{(k_C-1)!(k-k_C)!}q_{C|C}^{k_C-1}q_{D|C}^{k-k_C}+O(w^2)
  \nonumber \\  & &
  +p_Cq_{C|C}q_{D|C}\frac{k-1}{k}(\pi_D^C-\pi_C^C)w\sum_{k_C=1}^{k-1}\frac{(k-2)!}{(k_C-1)!(k-k_C-1)!}q_{C|C}^{k_C-1}q_{D|C}^{k-k_C-1}
    \nonumber \\  &=&
     p_C-p_Cq_{C|C}+p_Cq_{C|C}q_{D|C}\frac{k-1}{k}(\pi_D^C-\pi_C^C)w+O(w^2)
   \nonumber \\  &=&
     p_{CD}+\frac{k-1}{k}(\pi_D^C-\pi_C^C)p_{CD}q_{C|C}w+O(w^2).
\end{eqnarray}

Substituting equations~(\ref{pcposi}) and~(\ref{pcneg}) into equation~(\ref{pc}), we obtain
\begin{eqnarray}
  \dot{p_C}
  &=&    \frac{1}{N}\frac{k-1}{k}p_{CD}\left[q_{D|D}(\pi_C^D-\pi_D^D)+q_{C|C}(\pi_C^C-\pi_D^C)\right]w+O(w^2).
  \label{dotpc}
\end{eqnarray}

The first term in the right-hand side of the equation~(\ref{pcc}) is
\begin{eqnarray*}
  \sum_{k_C=0}^{k}\frac{2k_C}{kN}\mathrm{Pr}\left(\Delta p_{CC}=\frac{2k_C}{kN}\right)
&=&
    \sum_{k_C=0}^{k}\frac{2k_C}{kN}p_D{k \choose k_C}q_{C|D}^{k_C}q_{D|D}^{k_D}\left(\frac{k_C}{k}+O(w)\right)
\nonumber \\ &=&
    \frac{2p_D}{Nk^2}\sum_{k_C=0}^{k}{k \choose k_C}q_{C|D}^{k_C}q_{D|D}^{k_D}k_C^2+O(w)
\nonumber \\ &=&
   \frac{2p_D}{Nk^2}\sum_{k_C=1}^{k}\frac{k_C^2k!}{k_C!(k-k_C)!}q_{C|D}^{k_C}q_{D|D}^{k-k_C}+O(w)
\nonumber \\ &=&
   \frac{2p_D}{Nk^2}\sum_{k_C=1}^{k}\frac{q_{C|D}(k_C-1+1)k!}{(k_C-1)!(k-k_C)!}q_{C|D}^{k_C-1}q_{D|D}^{k-k_C}+O(w)
\nonumber \\ &=&
   \frac{2p_{CD}}{Nk^2}\bigg[\sum_{k_C=1}^{k}\frac{(k_C-1)k!}{(k_C-1)!(k-k_C)!}q_{C|D}^{k_C-1}q_{D|D}^{k-k_C}
\nonumber \\ & &
   +\sum_{k_C=1}^{k}\frac{k(k-1)!}{(k_C-1)!(k-k_C)!}q_{C|D}^{k_C-1}q_{D|D}^{k-k_C}\bigg]+O(w)
\nonumber \\ &=&
    \frac{2p_{CD}}{Nk^2}\left[\sum_{k_C=2}^{k}\frac{q_{C|D}k(k-1)(k-2)!}{(k_C-2)!(k-k_C)!}q_{C|D}^{k_C-2}q_{D|D}^{k-k_C}
   +k\right]+O(w)
\nonumber \\ &=&
    \frac{2p_{CD}}{Nk^2}\left[q_{C|D}k(k-1)+k\right] +O(w)
\nonumber \\ &=&
    \frac{2p_{CD}}{Nk}\left[1+(k-1)q_{C|D}\right]+O(w),
\end{eqnarray*}
and the second term is
\begin{eqnarray*}
  \sum_{k_C=0}^{k}\left(-\frac{2k_C}{kN}\right)\mathrm{Pr}\left(\Delta p_{CC} = -\frac{2k_C}{kN}\right)
&=&
  -\sum_{k_C=0}^{k}\frac{2k_C}{kN}p_C{k \choose k_C}q_{C|C}^{k_C}q_{D|C}^{k_D}\left( \frac{k_D}{k}+O(w)\right)
\nonumber \\ &=&
    -\frac{2p_C}{Nk^2}\sum_{k_C=0}^{k}{k \choose k_C}q_{C|C}^{k_C}q_{D|C}^{k_D}k_Ck_D+O(w)
\nonumber \\ &=&
    -\frac{2p_C}{Nk^2}\sum_{k_C=1}^{k-1}\frac{k!}{k_C!(k-k_C)!}q_{C|C}^{k_C}q_{D|C}^{k-k_C}k_C(k-k_C)+O(w)
\nonumber \\ &=&
    -\frac{2p_C}{Nk^2}\sum_{k_C=1}^{k-1}\frac{q_{C|C}q_{D|C}k(k-1)(k-2)!}{(k_C-1)!(k-k_C-1)!}q_{C|C}^{k_C-1}q_{D|C}^{k-k_C-1}+O(w)
\nonumber \\ &=&
    -\frac{2(k-1)}{Nk}p_{CD}q_{C|C}+O(w).
\end{eqnarray*}
Hence, we obtain
\begin{eqnarray}
  \dot{p}_{CC}
  &=&
    \frac{2}{Nk}p_{CD}\left[1+(k-1)(q_{C|D}-q_{C|C})\right]+O(w).
  \label{dotpcc}
\end{eqnarray}

From equations (\ref{dotpc}) and (\ref{dotpcc}), we have
\begin{eqnarray}
  \dot{q}_{C|C} &=& \frac{\mathrm{d}}{\mathrm{d}t}\left(\frac{p_{CC}}{p_C}\right)
   \nonumber \\ &=&  \frac{2}{Nk}\frac{p_{CD}}{p_C}\left[1+(k-1)(q_{C|D}-q_{C|C})\right]+O(w).
\label{qcc}
\end{eqnarray}

The system is described only by $p_C$ and $q_{C|C}$.
Rewriting the r.h.s's of equations~(\ref{dotpc}) and (\ref{qcc}) as functions of $p_C$ and $q_{C|C}$ yields the closed dynamic system
\begin{eqnarray}
\label{system}
 \left\{
 \begin{array}{cccc}
  \dot{p}_C &=& wF_1\left(p_C,q_{C|C}\right)+O(w^2) \\
  \dot{q}_{C|C} &=& F_2\left(p_C,q_{C|C}\right)+O(w)
  \end{array}
  \right.
\end{eqnarray}
where
\begin{eqnarray*}
F_1\left(p_C,q_{C|C}\right) &=&
\frac{1}{N}\frac{k-1}{k}p_C(1-q_{C|C})\Big[q_{C|C}(\pi_C^C-\pi_D^C)
\nonumber \\ & &
+\frac{1-2p_C+p_Cq_{C|C}}{1-p_C}(\pi_C^D-\pi_D^D)\Big],
\nonumber \\
F_2\left(p_C,q_{C|C}\right) &=&
\frac{2}{Nk}(1-q_{C|C}) \left[1+(k-1)\frac{p_C-q_{C|C}}{1-p_C}\right].
\end{eqnarray*}
With $0<w\ll1$, above system can be reduced as
\begin{eqnarray*}
 \left\{
 \begin{array}{cccc}
  \dot{p}_C &=& wF_1\left(p_C,q_{C|C}\right) \\
  w\dot{q}_{C|C} &=& wF_2\left(p_C,q_{C|C}\right)
  \end{array}
  \right..
\end{eqnarray*}
For $q_{C|C}$, whose velocity can be large when $w$ is small and $wF_2(p_C,q_{C|C})\neq0$, may rapidly converge to the root defined by $F_2(p_C,q_{C|C})=0$ as time $t\rightarrow+\infty$. Thus we get
\begin{equation}
\label{basic21}
  q_{C|C}=\frac{k-2}{k-1}p_C+\frac{1}{k-1}
\end{equation}
and the reduced (slow) model
\begin{eqnarray*}
  \dot{p}_C &=& wF_1\left(p_C,\frac{k-2}{k-1}p_C+\frac{1}{k-1}\right).
\end{eqnarray*}

From equation~(\ref{basic21}), we know that all variables in the dynamically evolutionary system (\ref{system}) can be described only by $p_C$ when it is stable.
And from equations~(\ref{basic1})$\sim$(\ref{basic5}), we have
\begin{eqnarray}
 \label{basic222}
  q_{D|C} &=& \frac{k-2}{k-1}(1-p_C), \\
   \label{basic22}
  p_{CD} &=& p_{DC} = \frac{k-2}{k-1}p_C(1-p_C), \\
  \label{basic23}
  q_{C|D} &=& \frac{k-2}{k-1}p_C, \\
  \label{basic24}
  q_{D|D} &=& 1-\frac{k-2}{k-1}p_C, \\
  \label{basic25}
  p_{DD} &=& (1-p_C)\left(1-\frac{k-2}{k-1}p_C\right).
\end{eqnarray}
Here we obtain
\begin{eqnarray}
\label{appendix_nete1}
  q_{C|C}-q_{C|D}&=&\frac{1}{k-1} \\
\label{appendix_nete2}
  q_{D|D}-q_{D|C}&=&\frac{1}{k-1}
\end{eqnarray}
   as the evolution is stable.
That is to say, in finite structured population where every player has $k$ neighbors, there is a high average probability for cooperators or defectors to form clusters where every individual with the same strategy.
We know that the above equation is irrelevant to the payoff forms and the existence of nonlinearity in individual fitness.
The relation is also proposed in \cite{Ohtsuki06Nature} for pairwise interactions.
Hence, it explicitly illustrates the fact of clustering by local interaction for multiple and pairwise games in finite structured population.

Then, as $\delta\neq1$, we know
\begin{eqnarray}
  \pi_C^D-\pi_D^D
  &=&
  P_C\left((k-1)q_{C|C}+1\right)+(k-1)q_{C|C}\cdot P_C\left((k-1)q_{C|C}+2\right)
 \nonumber \\ & &
 +(k-1)q_{D|C}\cdot P_C\left((k-1)q_{C|D}+1\right)
 \nonumber \\ & &
 -P_D\left((k-1)q_{C|D}\right)-(k-1)q_{C|D}\cdot P_D\left((k-1)q_{C|C}+1\right)
  \nonumber \\ & &
 -(k-1)q_{D|D}\cdot P_D\left((k-1)q_{C|D}\right)-c
\nonumber \\ &=&
\frac{b}{(k+1)(1-\delta)}\bigg[1-\delta^{(k-1)q_{C|C}+1}+(k-1)q_{C|C}
-(k-1)q_{C|C}\delta^{(k-1)q_{C|C}+2}
\nonumber \\ & &
+(k-1)q_{D|C}-(k-1)q_{D|C}\delta^{(k-1)q_{C|D}+1}-1
+\delta^{(k-1)q_{C|D}}-(k-1)q_{C|D}
\nonumber \\ & &
+(k-1)q_{C|D}\delta^{(k-1)q_{C|C}+1}-(k-1)q_{D|D}
+(k-1)q_{D|D}\delta^{(k-1)q_{C|D}}\bigg]-(k+1)c
\nonumber \\ &=&
\frac{b}{(k+1)(1-\delta)}
\bigg\{-(k-1)q_{C|C}\delta^{(k-1)q_{C|C}+2}+\big[(k-1)q_{C|D}-1\big]\delta^{(k-1)q_{C|C}+1}  \nonumber \\ & &
-(k-1)q_{D|C}\delta^{(k-1)q_{C|D}+1}+\big[(k-1)q_{D|D}+1\big]\delta^{(k-1)q_{C|D}}\bigg\}
-(k+1)c
\nonumber \\ &=&
\frac{b}{(k+1)(1-\delta)}
\bigg\{\big[-(k-2)p_C-1\big]\delta^{(k-2)p_C+3}+\big[(k-2)p_C-1\big]\delta^{(k-2)p_C+2}
\nonumber \\ & &
+\big[(k-2)p_C-k+2\big]\delta^{(k-2)p_C+1}+\big[-(k-2)p_C+k\big]\delta^{(k-2)p_C}  \bigg\}-(k+1)c
\nonumber \\ &=&
\frac{b}{(k+1)(1-\delta)}
\bigg[-(k-2)p_C\delta^{(k-2)p_C+3}+(k-2)p_C\delta^{(k-2)p_C+2}+(k-2)p_C\delta^{(k-2)p_C+1}
\nonumber \\ & &
-(k-2)p_C\delta^{(k-2)p_C}
-\delta^{(k-2)p_C+3}-\delta^{(k-2)p_C+2}+2\delta^{(k-2)p_C+1}-k\delta^{(k-2)p_C+1}
\nonumber \\ & &
+k\delta^{(k-2)p_C}\bigg]-(k+1)c
\nonumber \\ &=&
\frac{b(k-2)(\delta^2-1)}{k+1}\delta^{(k-2)p_C}p_C
+\frac{b(\delta^2+2\delta+k)}{k+1}\delta^{(k-2)p_C}-(k+1)c.
\label{pid}
\end{eqnarray}
Using
\begin{eqnarray*}
  \pi_C^D-\pi_D^D-\left(\pi_C^C-\pi_D^C\right) &=&
P_C\left((k-1)q_{C|C}+1\right)-P_C\left((k-1)q_{C|C}+2\right)
\nonumber \\ & &
+P_D\left((k-1)q_{C|D}+1\right)-P_D\left((k-1)q_{C|D}\right)
\nonumber \\ &=&
\frac{b(1-\delta^2)}{k+1}\delta^{(k-2)p_C},
\end{eqnarray*}
we get
\begin{eqnarray}
  \pi_C^C-\pi_D^C &=&
\frac{b(k-2)(\delta^2-1)}{k+1}\delta^{(k-2)p_C}p_C
+\frac{b\big[2\delta^2+2\delta+k-1\big]}{k+1}\delta^{(k-2)p_C}-(k+1)c.
\label{pic}
\end{eqnarray}

As $\delta=1$, we obtain
\begin{eqnarray*}
 \pi^D_C-\pi^D_D &=&  \pi^C_C-\pi^C_D =
 \frac{b(k+3)}{k+1}-(k+1)c.
\label{pid1}
\end{eqnarray*}
From above equation, we find both equation~(\ref{pid}) and (\ref{pic}) are available as $\delta=1$.

\subsection{Fixation probability}
\label{appendixe}
Using Kolmogorov backward equation \cite{Ewens2004a,CrowandKimura2009,Karlin1981a},
we get the fixation probability, $\phi_C(p)$ of the strategy cooperation with initial frequency $p$, satisfies the ordinary differential equation as
\begin{equation*}
  M\left(p_C\right)\frac{\mathrm{d}\phi_C(p)}{\mathrm{d}p}
  +\frac{V\left(p_C\right)}{2}\frac{\mathrm{d}^2\phi_C(p)}{\mathrm{d}p^2}=0,
\end{equation*}
with boundary conditions
\begin{eqnarray*}
  \phi_C(0)=0 &~~~~& \phi_C(1)=1
\end{eqnarray*}
where $M\left(p_C\right)$ and $V\left(p_C\right)$ are respectively the mean and the variance of $p_C$, the amount of change in strategy frequency per generation.
The solution of the above differential equation with the boundary conditions can be written
\begin{equation*}
  \phi_C(p)=\frac{\int_0^pG(x)\,\mathrm{d}x}{\int_0^1 G(x)\,\mathrm{d}x}
\end{equation*}
where
\begin{equation*}
  G(x)=\mathrm{e}^{-\int\frac{2M(x)}{V(x)}\mathrm{d}x}.
\end{equation*}

Now, let's calculate $-2M(x)/V(x)$.
In a short time interval, $\Delta t$, we have
\begin{eqnarray}
 M(p_C) &=& \frac{\mathrm{E}[\Delta p_C]}{\Delta t}  =
  \frac{\Delta t}{N}\left[\mathrm{Pr}\left(\Delta p_C=\frac{1}{N}\right)-\mathrm{Pr}\left(\Delta p_C=-\frac{1}{N}\right)\right]\frac{1}{\Delta t}
\nonumber \\  &\approx&  \frac{1}{N}\frac{k-1}{k}p_{CD}\Big[q_{D|D}(\pi_C^D-\pi_D^D)+q_{C|C}(\pi_C^C-\pi_D^C)\Big]w
\nonumber \\ &=&
  \frac{w}{N}\frac{k-1}{k}\frac{k-2}{k-1}p_C(1-p_C)
  \left[\left(\frac{k}{k-1}-q_{C|C}\right)(\pi^D_C-\pi^D_D)+q_{C|C}(\pi^C_C-\pi^C_D)\right]
\nonumber \\ &=&
 \frac{w(k-2)p_C(1-p_C)}{Nk(k-1)}
  \bigg\{k(\pi^D_C-\pi^D_D)+\left[(k-2)p_C+1\right]\left[(\pi^C_C-\pi^C_D)-(\pi^D_C-\pi^D_D) \right]\bigg\}
\nonumber \\ &=&
 \frac{w(k-2)p_C(1-p_C)}{Nk(k-1)}\Bigg\{k\bigg[\frac{b(k-2)(\delta^2-1)}{k+1}\delta^{(k-2)p_C}p_C
+\frac{b(\delta^2+2\delta+k)}{k+1}\delta^{(k-2)p_C}
\nonumber \\ & &
-(k+1)c\bigg]  +\left[(k-2)p_C+1\right]\frac{b(\delta^2-1)\delta^{(k-2)p_C}}{k+1}\Bigg\}
\nonumber \\ &=&
 \frac{w(k-2)p_C(1-p_C)}{Nk(k-1)}\Bigg\{b(\delta^2-1)(k-2)\delta^{(k-2)p_C}p_C
\nonumber \\ & &
+\frac{b\big[(k+1)\delta^2+2k\delta+k^2-1\big]}{k+1}\delta^{(k-2)p_C}
-k(k+1)c\Bigg\},
\label{mpc}
\end{eqnarray}
and
\begin{eqnarray}
   V(p_C) &=& \frac{\mathrm{Var}[\Delta p_C]}{\Delta t}  =
  \frac{1}{\Delta t}\mathrm{E}\left[\left(\Delta p_C-\frac{\mathrm{E}[\Delta p_C]}{\Delta t}\right)^2\right]\Delta t
\nonumber \\ &=&
  \left[\frac{1}{N^2}-2\frac{\mathrm{E}[\Delta p_C]}{N\Delta t}+\left(\frac{\mathrm{E}[\Delta p_C]}{\Delta t}\right)^2\right]\mathrm{Pr}\left(\Delta p_C=\frac{1}{N}\right)
\nonumber \\ & &
  +\left[\frac{1}{N^2}+2\frac{\mathrm{E}[\Delta p_C]}{N\Delta t}+\left(\frac{\mathrm{E}[\Delta p_C]}{\Delta t}\right)^2\right]\mathrm{Pr}\left(\Delta p_C=-\frac{1}{N}\right)
\nonumber \\ & \approx &
  \frac{1}{N^2}\left[\mathrm{Pr}\left(\Delta p_C=\frac{1}{N}\right)+\mathrm{Pr}\left(\Delta p_C=-\frac{1}{N}\right)\right]
\nonumber \\ & \approx &
  \frac{2p_{CD}}{N^2}  =
   \frac{2p_C(k-2)(1-p_C)}{N^2(k-1)}.
\label{vpc}
\end{eqnarray}

Then we obtain
\begin{eqnarray*}
  -\frac{2M(x)}{V(x)} &=& -\frac{Nw}{k}\Bigg\{b(\delta^2-1)(k-2)\delta^{(k-2)x}x
\nonumber \\ & &
+\frac{b\big[(k+1)\delta^2+2k\delta+k^2-1\big]}{k+1}\delta^{(k-2)x}
-k(k+1)c\Bigg\}
\nonumber \\ &=&
w\Bigg\{-\frac{Nb(\delta^2-1)(k-2)(k+1)}{k(k+1)}\delta^{(k-2)x}x
\nonumber \\ & &
-\frac{Nb\big[(k+1)\delta^2+2k\delta+k^2-1\big]}{k(k+1)}\delta^{(k-2)x}
+N(k+1)c\Bigg\}.
\end{eqnarray*}
Therefore, we have
\begin{eqnarray*}
\label{2mv}
  -\frac{2M(x)}{V(x)} &=&
w\Big[R\delta^{(k-2)x}x+S\delta^{(k-2)x}+N(k+1)c\Big]
\end{eqnarray*}
where
\begin{eqnarray*}
\label{notationsrs}
  R = -\frac{Nb(\delta^2-1)(k-2)}{k}
\quad
  S = -\frac{Nb\big[(k+1)\delta^2+2k\delta+k^2-1\big]}{k(k+1)}.
\end{eqnarray*}

\subsubsection{Fixation probability for $\delta=1$}
As $\delta=1$, $k\geq2$, we have
\begin{eqnarray*}
R=0, \quad S=-\frac{Nb(k+3)}{k+1}, \quad \text{and} -\frac{2M(x)}{V(x)}=[S+c(k+1)N]w.
\end{eqnarray*}
Then we get
\begin{equation*}
  G(x) =\mathrm{e}^{\int\left[S+c(k+1)N\right]w\mathrm{d}x}
  \approx \Big\{1+\left[S+c(k+1)N\right]wx\Big\}C_0
\end{equation*}
where $C_0$ is a constant.
Thus,
\begin{eqnarray*}
  \phi_C(p)&=&\frac{p+\left[S+c(k+1)N\right]\frac{p^2}{2}w}{1+\left[S+c(k+1)N\right]\frac{1}{2}w}
\nonumber \\ &\approx&
 p+\Big[S+c(k+1)N\Big]\frac{p^2}{2}w-\Big[S+c(k+1)N\Big]\frac{p}{2}w
\nonumber \\ &=&
 p+\Big[S+c(k+1)N\Big]\frac{w}{2}(p^2-p)
\nonumber \\ &=&
p+\frac{N}{2}\left[\frac{b(k+3)}{k+1}-c(k+1)\right]p(1-p)w.
\end{eqnarray*}
And as $\delta=1$, $k\geq2$, we also have
\begin{eqnarray*}
  \phi_D(p)&=&
p-\frac{N}{2}\left[\frac{b(k+3)}{k+1}-c(k+1)\right]p(1-p)w.
\end{eqnarray*}

\subsubsection{Fixation probability for $\delta\neq1$ and $k\neq2$}
As $\delta\neq1$, and $k\neq2$, we have
\begin{eqnarray*}
  \int-\frac{2M(x)}{V(x)}\,\mathrm{d}x &=& \int\left[(Rx+S)\delta^{(k-2)x}+c(k+1)N\right]w\,\mathrm{d}x
\nonumber \\ &=&
w\left[(Rx+S)\frac{\delta^{(k-2)x}}{(k-2)\ln\delta}-\int R\frac{\delta^{(k-2)x}}{(k-2)\ln\delta}\mathrm{d}x+c(k+1)Nx+C_2\right]
\nonumber \\ &=&
  c(k+1)Nxw+\frac{\delta^{(k-2)x}}{(k-2)\ln\delta}\left[Rx+S-\frac{R}{(k-2)\ln\delta}\right]w+C_3w
\end{eqnarray*}
where $C_2$ and $C_3$ are constants, and then we obtain
\begin{eqnarray*}
  G(x) &=& \mathrm{e}^{\left\{c(k+1)Nx+\frac{\delta^{(k-2)x}}{(k-2)\ln\delta}
  \left[Rx+S-\frac{R}{(k-2)\ln\delta}\right]+C_3\right\}w}
\nonumber \\ &\approx&
  1+\left\{c(k+1)Nx+\frac{\delta^{(k-2)x}}{(k-2)\ln\delta}
  \left[Rx+S-\frac{R}{(k-2)\ln\delta}\right]+C_3\right\}w.
\end{eqnarray*}
Let $A=(k-2)\ln\delta$, then
\begin{eqnarray*}
  \int_0^pG(x)\,\mathrm{d}x &\approx&
  x\Big|^p_0+\left[\frac{c(k+1)Nx^2}{2}\bigg|^p_0+\frac{\delta^{(k-2)x}\left(Rx+S-\frac{R}{A}\right)}{A^2}\bigg|^p_0-\frac{R\delta^{(k-2)x}}{A^3}\bigg|^p_0   +C_3\bigg|^p_0\right]w
 \nonumber \\ &=&
   p+\left[\frac{c(k+1)Np^2}{2}+\frac{1}{A^2}\delta^{(k-2)p}\left(Rp+S-\frac{R}{A}\right)
  -\frac{1}{A^2}\left(S-\frac{R}{A}\right)
  -\frac{R\delta^{(k-2)p}}{A^3}+\frac{R}{A^3}+C_3p\right]w
   \nonumber \\ &=&
   p+\left[\frac{c(k+1)Np^2}{2}+\left(\frac{R}{A^2}p+\frac{S}{A^2}-\frac{2R}{A^3}\right)\delta^{(k-2)p}-\left(\frac{S}{A^2}-\frac{2R}{A^3} \right) +C_3p  \right]w
\end{eqnarray*}
and
\begin{equation*}
  \int_0^1G(x)\,\mathrm{d}x \approx
  1+\left[\frac{c(k+1)N}{2}+\left(\frac{R}{A^2}+\frac{S}{A^2}-\frac{2R}{A^3}\right)\delta^{(k-2)}-\left(\frac{S}{A^2}-\frac{2R}{A^3} \right) +C_3   \right]w.
\end{equation*}

Thus, the fixation probability of the strategy cooperation, $\phi_C(p)$, with initial frequency $p_C(t=0)=p$ can be written
\begin{equation*}
  \phi_C(p)=\frac{A_1+B_1w}{C_1+D_1w}
\end{equation*}
where
\begin{eqnarray}
  A_1=p &~~~~~& B_1=\frac{c(k+1)Np^2}{2}+\left(\frac{R}{A^2}p+\frac{S}{A^2}-\frac{2R}{A^3}\right)\delta^{(k-2)p}-\left(\frac{S}{A^2}-\frac{2R}{A^3} \right)+C_3p,
   \nonumber \\
 C_1=1 &~~~~~& D_1=\frac{c(k+1)N}{2}+\left(\frac{R}{A^2}+\frac{S}{A^2}-\frac{2R}{A^3}\right)\delta^{(k-2)}-\left(\frac{S}{A^2}-\frac{2R}{A^3} \right)+C_3.
  \nonumber
\end{eqnarray}
Expanding $\phi_C(p)$ in a Taylor series at $w=0$, we obtain
\begin{eqnarray*}
  \phi_C(p) &\approx& p+(B_1-D_1p)w
\nonumber \\ &=&
  p+\left[\left(\frac{c(k+1)N}{2}p+Q\right)(p-1)
  +\left(\frac{R}{A^2}p+Q\right)\delta^{(k-2)p}
  -\left(\frac{R}{A^2}+Q\right)p\delta^{k-2}\right]w
\end{eqnarray*}
where
\begin{eqnarray*}
  Q &=& \frac{S}{A^2}-\frac{2R}{A^3}
=
  \frac{-ANb\left[(k+1)\delta^2+2k\delta+k^2-1\right]+2Nb(\delta^2-1)(k-2)(k+1)}{A^3k(k+1)}
\nonumber \\ &=&
  Nb\frac{2(\delta^2-1)(k-2)(k+1)-A\left[(k+1)\delta^2+2k\delta+k^2-1\right]}{A^3k(k+1)}
  = Nb\frac{Q_2}{Q_1},\\
\frac{R}{A^2}p+Q &=& -\frac{pNb(\delta^2-1)(k-2)(k+1)}{A^2k(k+1)}
+\frac{NbQ_2}{A^3k(k+1)}
= \frac{ApNb(1-\delta^2)(k-2)(k+1)+NbQ_2}{A^3k(k+1)}
\nonumber \\ &=&
  Nb\frac{pQ_3+Q_2}{Q_1}, \\
\frac{R}{A^2}+Q &=& Nb\frac{Q_3+Q_2}{Q_1},
\end{eqnarray*}
and the notations are used as
\begin{eqnarray*}
 \label{notationsq1}
  Q_1 &=& A^3k(k+1) = k(k+1)(k-2)^3\ln^3\delta, \\
  Q_2 &=& 2(\delta^2-1)(k-2)(k+1)-A\left[(k+1)\delta^2+2k\delta+k^2-1\right]  \nonumber
  \\ &=&
 2(\delta^2-1)(k-2)(k+1)-\left[(k-2)(k+1)\delta^2+2k(k-2)\delta+(k-2)(k^2-1)\right]\ln\delta,
  \label{notationsq2}
   \\
  Q_3 &=& A(1-\delta^2)(k-2)(k+1) = (1-\delta^2)(k+1)(k-2)^2\ln\delta.
    \label{notationsq3}
\end{eqnarray*}

Furthermore we have
\begin{eqnarray*}
  \phi_C(p) &\approx&
  p+\left\{\left[\frac{c(k+1)N}{2}p+Nb\frac{Q_2}{Q_1}\right](p-1)
  +Nb\frac{pQ_3+Q_2}{Q_1}\delta^{(k-2)p}
  -Nb\frac{Q_2+Q_3}{Q_1}p\delta^{k-2}\right\}w
\end{eqnarray*}
with $\delta\neq1$, and $k\neq2$.

And we also have
\begin{eqnarray*}
\label{phidp01}
  \phi_D(p) &\approx&
  p-\left\{\left[\frac{c(k+1)N}{2}p+Nb\frac{Q_2+Q_3}{Q_1}\delta^{k-2}\right](p-1)
  -Nb\frac{(p-1)Q_3-Q_2}{Q_1}\delta^{(k-2)(1-p)}
  -Nbp\frac{Q_2}{Q_1}\right\}w
\end{eqnarray*}
with $\delta\neq1$, and $k\neq2$.

\subsubsection{Fixation probability for $k=2$}
Using the same method as above, we have
\begin{eqnarray*}
  \phi_C(p) &\approx&
  p+\frac{N}{2}\left[\frac{(3\delta^2+4\delta+3)b}{6}-3c\right]p(1-p)w
\end{eqnarray*}
with $k=2$,
and
\begin{eqnarray*}
  \phi_D(p) &\approx&
  p-\frac{N}{2}\left[\frac{(3\delta^2+4\delta+3)b}{6}-3c\right]p(1-p)w
\end{eqnarray*}
with $k=2$.

\subsubsection{Conclusion of the fixation probability for cooperation and defection}
By the above tedious calculation, we obtain the fixation probability of strategy cooperation with initial frequency $p$ as
\begin{equation}
 \phi_C(p) =
 \begin{cases}
 p+\frac{N}{2}\left[\frac{b(k+3)}{k+1}-c(k+1)\right]p(1-p)w
       & \text{for $\delta=1$}  \\
 p+\frac{N}{2}\left[\frac{(3\delta^2+4\delta+3)b}{6}-3c\right]p(1-p)w
      & \text{for $k=2$}  \\
  p+\left\{\left[\frac{c(k+1)N}{2}p+Nb\frac{Q_2}{Q_1}\right](p-1)
  +Nb\frac{pQ_3+Q_2}{Q_1}\delta^{(k-2)p}
  -Nb\frac{Q_2+Q_3}{Q_1}p\delta^{k-2}\right\}w
      & \text{otherwise}
 \end{cases}.
\end{equation}
And through similarly calculative process, we have  the fixation probability of strategy defection with initial frequency $p$ as
\begin{equation}
 \phi_D(p) =
 \begin{cases}
 p-\frac{N}{2}\left[\frac{b(k+3)}{k+1}-c(k+1)\right]p(1-p)w
       & \text{for $\delta=1$}  \\
 p-\frac{N}{2}\left[\frac{(3\delta^2+4\delta+3)b}{6}-3c\right]p(1-p)w
      & \text{for $k=2$}  \\
  p-\left\{\left[\frac{c(k+1)N}{2}p+Nb\frac{Q_2+Q_3}{Q_1}\delta^{k-2}\right](p-1)
  -Nb\frac{(p-1)Q_3-Q_2}{Q_1}\delta^{(k-2)(1-p)}
  -Nbp\frac{Q_2}{Q_1}\right\}w
      & \text{otherwise}
 \end{cases}.
\end{equation}

\section{Criteria for emergence and stabilization of cooperation}
\label{appendix_b}
As to the criteria for emergence and stabilization of cooperation, we compare (the fixation probability with which a cooperator will invade and take over a population of $N-1$ defectors) and $1/N$ (the fixation probability of a neutral mutant \cite{Nowak06book}).
If $\rho_C>1/N$ then natural selection favors cooperator replacing defector \cite{Nowak06book}
, we call that natural selection favors the emergence of cooperation.
We call that natural selection favors the stabilization of cooperation if $\rho_D<1/N$, that is, natural selection opposes the fixation of defectors.
And if $\rho_C>\rho_D$, we call that natural selection favors cooperator over defector \cite{Nowak06book}
.
For mathematical feasibility, we obtain the criteria according to the following three occasions.

\subsection{As $\delta=1$}
For large $N$ and $\delta=1$, the fixation probability of a single cooperator (defector) in a population of $N-1$ defectors (cooperators) is given by $\rho_C=\phi_C\left(\frac{1}{N}\right)$ ($\rho_D=\phi_D\left(\frac{1}{N}\right)$). That is,
\begin{eqnarray}
\label{rhocnodelta}
  \rho_C=\phi_C\left(\frac{1}{N}\right) &\approx& \frac{1}{N}+\frac{1}{2}\left[\frac{b(k+3)}{k+1}-c(k+1)\right]\left(1-\frac{1}{N}\right)w, \\
  \rho_D=\phi_D\left(\frac{1}{N}\right) &\approx& \frac{1}{N}-\frac{1}{2}\left[\frac{b(k+3)}{k+1}-c(k+1)\right]\left(1-\frac{1}{N}\right)w.
  \label{rhodnodelta}
\end{eqnarray}
Hence, we have $\rho_C>\frac{1}{N}$ if and only if
\begin{equation}
\label{last00a1}
  b/c>\frac{(k+1)^2}{k+3},
\end{equation}
and $\rho_D<\frac{1}{N}$ if and only if
\begin{equation}
\label{last00b1}
  b/c>\frac{(k+1)^2}{k+3}.
\end{equation}

\subsection{As $\delta\neq1$ and $k\neq2$}
For large $N$, $\delta\neq1$, and $k\neq2$, we have
\begin{eqnarray*}
\label{rhocdelta}
  \rho_C=\phi_C\left(\frac{1}{N}\right) &\approx& \frac{1}{N}+\frac{w}{Q_1}\left[
  \frac{1-N}{N}\frac{c(k+1)}{2}Q_1+(1-N)bQ_2+(Q_3+NQ_2)b\delta^{\frac{k-2}{N}}-(Q_3+Q_2)b\delta^{k-2}\right]
\nonumber \\ &=&
   \frac{1}{N}+\frac{w}{Q_1}\left[ \frac{1-N}{N}\frac{c(k+1)}{2}Q_1+bf(\delta) \right]
\end{eqnarray*}
where $f(\delta)=(NQ_2+Q_3)\delta^{\frac{k-2}{N}}-(Q_2+Q_3)\delta^{k-2}-(N-1)Q_2$.

As $0<\delta<1$, we know that $Q_1<0$, then
\begin{eqnarray*}
\rho_C>\frac{1}{N} &\Leftrightarrow&
  \frac{w}{Q_1}\left[ \frac{1-N}{N}\frac{c(k+1)}{2}Q_1+bf(\delta)\right]>0
\nonumber \\ &\Leftrightarrow&
\frac{1-N}{N}\frac{c(k+1)}{2}Q_1+bf(\delta)<0
\nonumber \\ &\Leftrightarrow&
\frac{bf(\delta)}{c}<\frac{(N-1)(k+1)Q_1}{2N}.
\end{eqnarray*}
We have $f(\delta)=(NQ_2+Q_3)\delta^{\frac{k-2}{N}}-(Q_2+Q_3)\delta^{k-2}-(N-1)Q_2<0$ (see Appendix \ref{appendix_d1}) as $0<\delta<1$.
Thus, we get
\begin{eqnarray*}
\rho_C>\frac{1}{N} &\Leftrightarrow&
b/c>\frac{(N-1)(k+1)Q_1}{2Nf(\delta)}
\end{eqnarray*}
for $0<\delta<1$.

As $\delta>1$, we know that $Q_1>0$, then
\begin{eqnarray*}
\rho_C>\frac{1}{N} &\Leftrightarrow&
  \frac{w}{Q_1}\left[ \frac{1-N}{N}\frac{c(k+1)}{2}Q_1+bf(\delta)\right]>0
\nonumber \\ &\Leftrightarrow&
\frac{1-N}{N}\frac{c(k+1)}{2}Q_1+bf(\delta)>0
\nonumber \\ &\Leftrightarrow&
\frac{bf(\delta)}{c}>\frac{(N-1)(k+1)Q_1}{2N}.
\end{eqnarray*}
We have $f(\delta)>0$ (see Appendix \ref{appendix_d2}) as $\delta>1$.
Thus, we get
\begin{eqnarray*}
\rho_C>\frac{1}{N} &\Leftrightarrow&
b/c>\frac{(N-1)(k+1)Q_1}{2Nf(\delta)}
\end{eqnarray*}
for $\delta>1$.

So, for large $N$, $\delta\neq1$, and $k\neq2$, we have $\rho_C>\frac{1}{N}$ if and only if
\begin{equation}
\label{last00a2}
b/c>\frac{(N-1)(k+1)Q_1}{2Nf(\delta)}
\end{equation}
where $f(\delta)=(NQ_2+Q_3)\delta^{\frac{k-2}{N}}-(Q_2+Q_3)\delta^{k-2}-(N-1)Q_2$.

For $\rho_D$, we have
\begin{eqnarray*}
\label{rhoddelta}
  \rho_D=\phi_D\left(\frac{1}{N}\right)
 &\approx&
   \frac{1}{N}-\frac{w}{Q_1}\bigg\{\frac{1-N}{N}\frac{c(k+1)}{2}Q_1-b(N-1)(Q_2+Q_3)\delta^{k-2}
\nonumber \\ & &
  +b\Big[NQ_2+(N-1)Q_3\Big]\delta^{\frac{(k-2)(N-1)}{N}}-bQ_2\bigg\}
\nonumber \\ &=&
   \frac{1}{N}-\frac{w}{Q_1}\left[\frac{1-N}{N}\frac{c(k+1)}{2}Q_1+bg(\delta)\right]
\end{eqnarray*}
where $g(\delta)=\left[NQ_2+(N-1)Q_3\right]\delta^{\frac{(k-2)(N-1)}{N}}-(N-1)(Q_2+Q_3)\delta^{k-2}-Q_2$.
We have $Q_1<0$, $g(\delta)<0$ as $0<\delta<1$, while $Q_1>0$, $g(\delta)>0$ as $\delta>1$.
Thus, using the same method, we also obtain $\rho_D<\frac{1}{N}$ if and only if
\begin{equation}
\label{last00b2}
b/c>\frac{(N-1)(k+1)Q_1}{2Ng(\delta)}
\end{equation}
for large $N$, $\delta\neq1$, and $k\neq2$.

\subsection{As $k=2$}
Using the same method as above, we have $\rho_C>\frac{1}{N}$, $\rho_D<\frac{1}{N}$  if and only if
\begin{equation}
\label{last00a3}
b/c>\frac{18}{3\delta^2+4\delta+3},
\end{equation}
and $\rho_D<\frac{1}{N}$  if and only if
\begin{equation}
\label{last00b3}
b/c>\frac{18}{3\delta^2+4\delta+3}
\end{equation}
for large $N$ and $k=2$.

\subsection{A rule for the evolution of cooperation}
\label{main_thesimplerule}
A rule is obtained theoretically for the evolution of cooperation under combination of structured population and multiple interactions, that is, we have proposition that $\rho_C>\frac{1}{N}$ if and only if
\begin{equation}
\label{phoccc}
 b/c>
 \begin{cases}
   \frac{(k+1)^2}{k+3}
      & \text{for $\delta=1$}  \\
  \frac{18}{3\delta^2+4\delta+3}
      & \text{for $k=2$}  \\
  \frac{(N-1)(k+1)Q_1}{2Nf(\delta)}
       & \text{otherwise}
 \end{cases},
\end{equation}
and $\rho_D<\frac{1}{N}$ if and only if
\begin{equation}
\label{phoddd}
 b/c>
 \begin{cases}
   \frac{(k+1)^2}{k+3}
      & \text{for $\delta=1$}  \\
  \frac{18}{3\delta^2+4\delta+3}
      & \text{for $k=2$}  \\
  \frac{(N-1)(k+1)Q_1}{2Ng(\delta)}
     & \text{otherwise}
 \end{cases}
\end{equation}
from inequalities \eqref{last00a1}, \eqref{last00a2}, \eqref{last00a3}, and \eqref{last00b1}, \eqref{last00b2}, \eqref{last00b3} for large $N$, $k>1$, and $\delta>0$,
where
\begin{eqnarray}
f(\delta)&=&(NQ_2+Q_3)\delta^{\frac{k-2}{N}}-(Q_2+Q_3)\delta^{k-2}-(N-1)Q_2, \\
g(\delta)&=&\Big[NQ_2+(N-1)Q_3\Big]\delta^{\frac{(k-2)(N-1)}{N}}-(N-1)(Q_2+Q_3)\delta^{k-2}-Q_2,
\end{eqnarray}
and
\begin{eqnarray*}
  Q_1 &=& k(k+1)(k-2)^3\ln^3\delta, \\
  Q_2 &=& 2(\delta^2-1)(k-2)(k+1)-\left[(k-2)(k+1)\delta^2+2k(k-2)\delta+(k-2)(k^2-1)\right]\ln\delta, \\
  Q_3 &=& (1-\delta^2)(k+1)(k-2)^2\ln\delta.
\end{eqnarray*}

For the linear public goods game on the cycle ($\delta=1$ and $k=2$), $\rho_C>\rho_D$ is equivalent to $b/c>9/5$,
which coincides with \cite{Veelen2012a} theoretically.
Through numerical simulations for $k>2$  to check our theoretical results, we find that the rule, $\rho_C>1/N$, is in good agreement with computer numerical simulations (see Fig.~\ref{fig_2}).

Furthermore, from the inequalities (\ref{phoccc}) and (\ref{phoddd}), we know that as $\delta \neq 1$ and $k\neq2$ the conditions for $\rho_C>\frac{1}{N}$  and $\rho_D<\frac{1}{N}$ are mainly different in $f(\delta)$ and $g(\delta)$.
Here we expend $f(\delta)$ and $g(\delta)$ in Taylor series around $\delta=1$ to investigate the difference.
For each $\delta$ in interval $(0.9,1.1)$, we have
\begin{eqnarray}
f(\delta)&=& \frac{k(k+3)(k-2)^3(N-1)}{2N}(\delta-1)^3 \nonumber \\ & &
+\frac{(4-3k^2)(N^2-3N+2)-k(19N^2-3N-16)+2k^3(N^2-1)}{12N^2}(k-2)^3(\delta-1)^4 \nonumber \\ & &
+o((\delta-1)^4) \\
g(\delta)&=& \frac{k(k+3)(k-2)^3(N-1)}{2N}(\delta-1)^3 \nonumber \\ & &
+\frac{(3k^2-4)(N^2-3N+2)-k(35N^2-51N+16)+k^3(4N^2-6N+2)}{12N^2}(k-2)^3(\delta-1)^4 \nonumber \\ & &
+o((\delta-1)^4)
\end{eqnarray}
where $o((\delta-1)^4)$ captures the error is the higher order infinitesimal of $(\delta-1)^4$.
Therefore we obtain $f(\delta)$ and $g(\delta)$ display the difference in forth order of $\delta-1$, and it is infinitesimal as $\delta$ around $1$.

\section{Theoretical analysis for the critical value of $b/c$ favoring evolution of cooperation}
\subsection{Evaluation on continuity for critical value of $b/c$}
\label{appendix_c1}
We first calculate the limitation of $\frac{(N-1)(k+1)Q_1}{2Nf(\delta)}$ and $\frac{(N-1)(k+1)Q_1}{2Ng(\delta)}$ at $\delta=1$ and $k\neq2$.
Let
\begin{eqnarray*}
\lim_{\delta\rightarrow1}\frac{(N-1)(k+1)Q_1}{2Nf(\delta)} &=& \frac{(N-1)(k+1)}{2N(L_1+L_2)}
\end{eqnarray*}
where
\begin{eqnarray}
L_1 &=&
\lim_{\delta\rightarrow1}\left[Q_3\left(\delta^{\frac{k-2}{N}}-\delta^{k-2}\right)\Big{/}Q_1\right]
\nonumber \\ &=&
\lim_{\delta\to1}\frac{(1-\delta^2)(\delta^{\frac{k-2}{N}}-\delta^{k-2})}{k(k-2)\ln^2\delta}
\nonumber \\ &=&
\lim_{\delta\to1}\frac{-2(\delta^{\frac{k-2}{N}+2}-\delta^{k})+(1-\delta^2)\left[\frac{k-2}{N}\delta^{\frac{k-2}{N}}-(k-2)\delta^{k-2}\right]}{2k(k-2)\ln\delta}
\nonumber \\ &=&
\lim_{\delta\to1}\frac{\delta}{2k(k-2)}\Bigg\{
-2\left[\left(\frac{k-2}{N}+2\right)\delta^{\frac{k-2}{N}+1}-k\delta^{k-1}\right]
-2\delta\left[\frac{k-2}{N}\delta^{\frac{k-2}{N}}-(k-2)\delta^{k-2}\right]
\nonumber \\ & &
+(1-\delta^2)\left[  \left(\frac{k-2}{N}\right)^2\delta^{\frac{k-2}{N}-1}-(k-2)^2\delta^{k-3}\right]\Bigg\}
\nonumber \\ &=&
\frac{2}{k}\left(1-\frac{1}{N}\right)
\end{eqnarray}
and
\begin{eqnarray}
L_2 &=&
\lim_{\delta\rightarrow1}\left\{Q_2\left[\left(\delta^{\frac{k-2}{N}}-\delta^{k-2}\right)+(N-1)\left(\delta^{\frac{k-2}{N}}-1\right)\right]\Big{/}Q_1\right\}
\nonumber \\ &=&
\lim_{\delta\to1}\frac{\overbrace{\left\{2(\delta^2-1)(k+1)-\left[(k+1)\delta^2+2k\delta+(k^2-1)\right]\ln\delta\right\}}^{L_3}
\overbrace{\left[(\delta^{\frac{k-2}{N}}-\delta^{k-2})+(N-1)(\delta^{\frac{k-2}{N}}-1)\right]}^{L_4}      }{\underbrace{k(k+1)(k-2)^2\ln^3\delta}_{L_5}}
\nonumber \\ &=&
\lim_{\delta\to1}\frac{L_3'L_4+L_3L_4'}{L_5'}    ~~~~~~~~~~~~\left(L_3(1)=L_4(1)=L_5(1)=L_5'(1)=0\right)
\nonumber \\ &=&
\lim_{\delta\to1}\frac{L_3''L_4+2L_3'L_4'+L_3L_4''}{L_5''}
~~~~~~~~~~~\left(L_4'(1)=L_5''(1)=0\right)
\nonumber \\ &=&
\lim_{\delta\to1}\frac{L_3'''L_4+3L_3''L_4'+3L_3'L_4''+L_3L_4'''}{L_5'''} ~~~~~~~~~~~\left(L_3'(1)\neq0,~L_4''(1)\neq0,~L_5'''\neq0\right)
\nonumber \\ &=&
\lim_{\delta\to1}\frac{3L_3'L_4''}{L_5'''}
 =
\frac{3(-k^2+k+4)(k-2)^2}{6k(k+1)(k-2)^2}\left(\frac{1}{N}-1\right)
\nonumber \\ &=&
\frac{-k^2+k+4}{2k(k+1)}\left(\frac{1}{N}-1\right)
\end{eqnarray}
with
\begin{eqnarray*}
  L_3' &=& 4(k+1)\delta-\left[2(k+1)\delta+2k\right]\ln\delta-\frac{(k+1)\delta^2+2k\delta+k^2-1}{\delta}, \\  L_4' &=&
   (k-2)(\delta^{\frac{k-2}{N}-1}-\delta^{k-3}), \\
     L_5' &=&
   \frac{3k(k+1)(k-2)^2\ln^2\delta}{\delta}, \\
    L_4'' &=&
   (k-2)\left[\left(\frac{k-2}{N}-1\right)\delta^{\frac{k-2}{N}-2}-(k-3)\delta^{k-4}\right], \\
       L_5'' &=&
   \frac{3k(k+1)(k-2)^2}{\delta^2}\left(2\ln\delta-\ln^2\delta\right), \\
    L_5'''(1) &=&
   \frac{6k(k+1)(k-2)^2}{\delta^3}.
\end{eqnarray*}
Therefore we get
\begin{equation}\label{infinitesimalrule}
\lim_{\delta\rightarrow1}\frac{(N-1)(k+1)Q_1}{2Nf(\delta)}=\frac{(k+1)^2}{k+3}.
\end{equation}
Using the same method, we also obtain
\begin{equation}\label{infinitesimalrule2}
\lim_{\delta\rightarrow1}\frac{(N-1)(k+1)Q_1}{2Ng(\delta)}=\frac{(k+1)^2}{k+3}.
\end{equation}
Thus, we find the greatest lower bounds which support $\rho_C>1/N$ and $\rho_D<1/N$ is continuous at $\delta=1$ respectively.
Also, we could obtain that the critical value is also continuous at $k=2$, or, $k \neq 2$ and $\delta \neq1$.
Thus, the critical values in equalities (\ref{phoccc}) and (\ref{phoddd}) can be proven as an continuous function of $k$ and $\delta$ for $k>1$ as well as $\delta>0$.

\subsection{The equivalent proposition with infinitesimal nonlinearity}
\label{appendix_c2}
For large structured population under linear public goods game or public goods game with nonlinearity in individual fitness on cycle, we have that natural selection favors emergence of cooperation if and only if it favors stabilization of cooperation (see equivalent proposition ($5$) in main text).
Here, we give the validation of the equivalent proposition for infinitesimal nonlinearity.

Expending the critical values in equalities (\ref{phoccc}) (indicated by $Critical~Value~C$) and (\ref{phoddd}) (indicated by $Critical~Value~D$) in Taylor series at $\delta=1$, we have
\begin{eqnarray*}
  Critical~Value~C &\approx& \frac{(k+1)^2}{k+3}
  -\frac{(1+k)^2\left[(N-2)(2+4k)+(N+1)(3k^2+k^3)\right]}{3Nk(3+k)^2}(\delta-1)
  \\ & &
  +\frac{(k+1)^2}{36N^2k^2(3+k)^3}\Big[16(N-2)^2+4(7N^2-46N+28)k+4(7N^2
  \\ & &
  -61N+1)k^2+2(53N^2-41N-40)k^3  +(65N^2+37N+5)k^4
  \\ & &
  +6(2N^2+6N+1)k^5+(N^2+5N+1)k^6 \Big] (\delta-1)^2
\end{eqnarray*}
\begin{eqnarray*}
  Critical~Value~D &\approx& \frac{(k+1)^2}{k+3}
  -\frac{(1+k)^2\left[(N-2)(-2-4k)+(2N-1)(3k^2+k^3)\right]}{3Nk(3+k)^2}(\delta-1)
    \\ & &
  +\frac{(k+1)^2}{36N^2k^2(3+k)^3}\Big[16(N-2)^2+4(11N^2+10N-28)k+4(-53N^2
  \\ & &
  +59N+1)k^2+2(-28N^2+121N-40)k^3 +(107N^2-47N+5)k^4
  \\ & &
  +6(9N^2-8N+1)k^5+(7N^2-7N+1)k^6 \Big] (\delta-1)^2.
\end{eqnarray*}
From above two equations, we find that the infinitesimal discounting effects inhibit evolution of cooperation, and the infinitesimal synergistic effects favor evolution of cooperation mostly for large population size $N$ comparing linear public goods game.

Besides, for large $N$, $\delta\rightarrow1$, and $k\geq2$, we have $\rho_C>\frac{1}{N}$ ($\rho_D<\frac{1}{N}$) if and only if
\begin{equation}
b/c>\frac{(k+1)^2}{k+3}.
\end{equation}
And for large structured population under public goods game with infinitesimal nonlinearity in individual fitness, we have
\begin{eqnarray}
  \rho_C>\frac{1}{N} \Leftrightarrow \rho_C>\rho_D  \Leftrightarrow \rho_D<\frac{1}{N}.
\end{eqnarray}

\section{Proof of the relation between $f(\delta)$ and $0$}
\label{appendix_d}
Let
\begin{eqnarray*}
  f(\delta) &=& \left(NQ_2+Q_3\right)\delta^{\frac{k-2}{N}}-\left(Q_2+Q_3\right)\delta^{k-2}-(N-1)Q_2
  \nonumber \\ &=&
  Q_2f_1(\delta)
  +f_2(\delta)
\end{eqnarray*}
where $f_1(\delta)=\delta^{\frac{k-2}{N}}-\delta^{k-2}+(N-1)\left(\delta^{\frac{k-2}{N}}-1\right)$, and $f_2(\delta)=Q_3\left(\delta^{\frac{k-2}{N}}-\delta^{k-2}\right)$.

For $Q_2$, we have
\begin{eqnarray*}
  Q_2 &=& 2(\delta^2-1)(k-2)(k+1)-\left[(k-2)(k+1)\delta^2+2k(k-2)\delta+(k-2)(k^2-1)\right]\ln\delta \\
  Q_2' &=& \frac{f_3(\delta)}{\delta}(k-2) \\
  Q_2'' &=& \left[1-2\ln\delta+\frac{k^2-2k\delta-1}{(k+1)\delta^2}\right](k+1)(k-2) \\
  Q_2''' &=& \frac{2\left[-k^2+k\delta(1-\delta)+1-\delta^2\right]}{\delta^3}(k-2)
\end{eqnarray*}
where $f_3(\delta)=(k+1)\left[(3-2\ln\delta)\delta^2-k+1\right]-2k\delta(\ln\delta+1)$.
Besides, $Q_1$, $Q_2$, $Q_3$, and the first to third derivative of $Q_2$ are all functions of $\delta$ with the parameter $k$.

\subsection{The proof of $f(\delta)<0$ when $0<\delta<1$}
\label{appendix_d1}
For large $N$ , $k\neq2$, and $0<\delta<1$, we derive $f(\delta)<0$ as follows.

(i). As $0<\delta<1$,  $f_2(\delta)<0$ since $Q_3<0$ and $\delta^{\frac{k-2}{N}}-\delta^{k-2}>0$.

(ii). $f_1(\delta)=N(\delta^{\frac{k-2}{N}}-1)+(1-\delta^{k-2})$. $0<1-\delta^{k-2}<1$ and $\delta^{\frac{k-2}{N}}-1<0$ for $0<\delta<1$. Hence, we have $f_1(\delta)<0$ when $N>\frac{1}{1-\delta^{\frac{k-2}{N}}}$.

(iii). As $0<\delta<1$, $Q_2''' <0$. $Q_2''(1)=k(k-1)(k-2)>0$, thus we have $Q_2''>0$ for $0<\delta<1$. With $Q_2'(1)=(-k^2+k+4)(k-2)<0$, we obtain that $Q_2'<0$ for $0<\delta<1$. And with $Q_2(1)=0$, we eventually get $Q_2>0$ for $0<\delta<1$.

 From (i)-(iii), we have $f(\delta)<0$ as $0<\delta<1$.

\subsection{The proof of $f(\delta)>0$ when $\delta>1$}
\label{appendix_d2}

For large $N$ , $k\neq2$, and $\delta>1$, we derive $f(\delta)>0$ as follows.

(i). As $\delta>1$,  $f_2(\delta)>0$ since $Q_3<0$ and $\delta^{\frac{k-2}{N}}-\delta^{k-2}<0$.

(ii). We have $f_1'(\delta)=(k-2)\left(\delta^{\frac{k-2}{N}-1}-\delta^{k-3}\right)<0$ when $\delta>1$. With $f_1(1)=0$, we know that $f_1(\delta)<0$ as $\delta>1$.

(iii). As $3\leq2\ln\delta$, \textit{i.e.}, $\delta\geq\mathrm{e}^{\frac{3}{2}}$ we have $Q_2<0$ since $Q_2'<0$ and $Q_2(1)=0$.

 (iv). Before proving $Q_2<0$ for $1<\delta<\mathrm{e}^{\frac{3}{2}}$, let us consider the function $h(x)=3(1+2x)-(3-2x)\mathrm{e}^{2x}$ first.
We want to point out as follows that $h(x)>0$ as $0<x<3/2$.

For $h(x)$, we have
\begin{eqnarray*}
h'(x)&=&4\mathrm{e}^{2x}(x-1)+6 \\
h''(x)&=&4\mathrm{e}^{2x}(2x-1)
\end{eqnarray*}
 Because $h''(x)<0$ as $0<x<1/2$,  and $h''(x)>0$ as $x>1/2$. we know that $h'(x)$ reaches its minimum at $x=1/2$ for $0<x<3/2$.
 Thus, we obtain $h'(x)>0$ for $0<x<3/2$ since $h'(1/2)=6-2\mathrm{e}>0$.
 And then we find $h(x)>0$ as $0<x<3/2$ since $h(0)=0$.

Let $\delta=\mathrm{e}^x$, and then $1<\delta<\mathrm{e}^{\frac{3}{2}}$ is equivalent to $0<x<\frac{3}{2}$.
 Hence, for $0<x<\frac{3}{2}$, we have
 \begin{eqnarray*}
f_3(\delta) &=& (k+1)\left[(3-2\ln\delta)\delta^2-k+1\right]-2k\delta(\ln\delta+1)
 \nonumber \\ &=&
 (k+1)\left[(3-2x)\mathrm{e}^{2x}-k+1\right]-2k\mathrm{e}^x(x+1)
 \nonumber \\ &<&
 (k+1)\left[3(1+2x)-k+1\right]-2k(1+x)(x+1)
  \nonumber \\ &=&
  -2kx^2+(2k+6)x-k^2+4+k
 \end{eqnarray*}
with the inequalities $3(1+2x)>(3-2x)\mathrm{e}^{2x}$ and $\mathrm{e}^x>1+x$.
The maximum value of $f_3(\delta)$ is $\overline{f_3(k)}=-k^2+\frac{3}{2}k+7+\frac{9}{2k}$ for $0<x<\frac{3}{2}$.
And $\overline{f_3(k)}<0$ if $k>3$.
It is meant to $Q_2<0$ with respect to $k>3$ and $1<\delta<\mathrm{e}^{\frac{3}{2}}$ since $Q'_2<0$  at this occasion.

 (v). For $k=3$ and $1<\delta<\mathrm{e}^{\frac{3}{2}}$, we have
  \begin{eqnarray*}
f_3'(\delta) &=& 16\delta(1-\ln\delta)-6\ln\delta-12 \\
f_3''(\delta) &=& -16\ln\delta-\frac{6}{\delta}
 \end{eqnarray*}
 and $f_3''(\delta)<0$.
 Because $f_3'(1)=4>0$ and $\underset{\delta\rightarrow+\infty}{\lim}f_3'=-\infty$, we derive that $f_3(\delta)$ reaches its maximum value at $\overline{\delta}$ where $f_3'(\overline{\delta})=0$.
 Through Matlab, we can get $f_3(\overline{\delta})<-0.9$.
 Hence, we obtain $Q_2<0$ as $k=3$ and $1<\delta<\mathrm{e}^{\frac{3}{2}}$ since $Q'_2<0$  at this occasion.

 From (iii)-(v), we have $Q_2<0$ as $\delta>1$ and $k\neq2$.

 Therefore, we have $f(\delta)>0$ as $\delta>1$ and $k\neq2$ from (i)-(v).


\end{document}